\documentclass[11pt,dvips]{article}
\usepackage{float}
\usepackage{graphicx}
\usepackage{amsmath}
\usepackage{amssymb}
\usepackage{latexsym}
\usepackage{color}
\providecommand{\openone}{\leavevmode\hbox{\small1\kern-4.3pt\normalsize1}}

\begin{document}
\thispagestyle{empty}
\begin{center}

\vspace{1.8cm}

%%%%%%%%%%%%%%%%%%%%%%%%%%%%%%%%%%%%%%%%%%%%%%%%%%%%%%%%%%%%%%%%%%%%%%%%%%%%%%%%%%%%%%%%%%%%%%%%%%%%%%%%%%%%%%%%%%%%%%%
   { \bf \Large The dynamic behaviors of local quantum uncertainty for three-qubit $X$ states under decoherence channels}\\
 %%%%%%%%%%%%%%%%%%%%%%%%%%%%%%%%%%%%%%%%%%%%%%%%%%%%%%%%%%%%%%%%%%%%%%%%%%%%%%%%%%%%%%%%%%%%%%%%%%%%%%%%%%%%%%%%%%%%%

\vspace{1.5cm}

{\bf A. Slaoui}$^{a,*}${\footnote { email: {\sf
*Corresponding author: abdallahsalaoui1992@gmail.com}}}, {\bf M. Daoud}$^{b,c}$ {\footnote {
email: {\sf m$_{-}$daoud@hotmail.com}}} and {\bf R. Ahl
Laamara}$^{a,d}$ {\footnote { email: {\sf ahllaamara@gmail.com}}}

\vspace{0.5cm}

$^{a}${\it LPHE-Modeling and Simulation, Faculty  of Sciences,
University
Mohammed V,\\ Rabat, Morocco.}\\[1em]

$^{b}${\it Department of Physics, Faculty of Sciences, University Ibn Tofail,\\ K\'enitra, Morocco.}\\[1em]

$^{c}${\it Abdus Salam International Centre for Theoretical Physics,\\ Miramare, Trieste, Italy.}\\[1em]

$^{d}${\it Centre of Physics and Mathematics,
CPM, CNESTEN,\\ Rabat, Morocco.}\\[1em]

\vspace{2cm} {\bf Abstract}
\end{center}
\baselineskip=18pt
\medskip

We derive the analytical expression of local quantum uncertainty for three qubit $X$-states. We give also the expressions of quantum discord and the negativity. A comparison of these three quantum correlations quantifiers is discussed in the special cases of mixed {\rm GHZ} states and Bell-type states. We find that local quantum uncertainty gives the same amount of non-classical correlations as are measured by entropic quantum discord and goes beyond negativity. We also discuss the dynamics of non-classical correlations under the effect of phase damping, depolarizing and phase reversal channels. We find the local quantum uncertainty shows more robustness and exhibits, under phase reversal effect, revival and frozen phenomena. The monogamy property of local quantum uncertainty is also discussed. It is shown that it is monogamous for three qubit states.\\

\newpage

%%%%%%%%%%%%%%%%%%%%%%%%%%%%%%%%%%%%%%%%%%%%%%%%%%%%%%%%%%%%%%%%%%%%%%%%%%%%%%%%%%%%%%%%%%%%%%%%%%%%%%%%
\section{Introduction}
%%%%%%%%%%%%%%%%%%%%%%%%%%%%%%%%%%%%%%%%%%%%%%%%%%%%%%%%%%%%%%%%%%%%%%%%%%%%%%%%%%%%%%%%%%%%%%%%%%%%%%%%
The characterization and quantification of quantum correlations in composite quantum systems is one of the most challenging topics in quantum information theory \cite{Nielsen,Bellac}. The interest in this field is motivated by the fact that entanglement can be used as useful resource for fundamental study in quantum mechanics and for applications in quantum teleportation \cite{Bouwmeester,Braunstein1998}, dense coding \cite{Mattle1996,Li2002} and quantum key distribution \cite{Bennett1992,Daoud2011}. The entanglement expresses the non local character of quantum mechanics theory \cite{Einstein1935,Bell1966}. To quantify the amount of entanglement, various measures have been proposed. The most familiar ones are the concurrence \cite{Yu2009,Wootters2001}, entanglement of formation \cite{BennettA1996,Popescu1997}, linear entropy \cite{Bose2000}, entanglement of distillation \cite{Bennett1996} and  the negativity \cite{Peres1996,Vidal2002}. Some results have shown that quantum correlations can not only be limited to entanglement, especially in mixed states, since separable quantum states can also have non-classical correlations. This yielded many works dedicated to introduce quantum correlation quantifiers beyond entanglement. In 2001, Ollivier and Zurek and independently Henderson and Vedral introduced the concept of entropic quantum discord as a quantifier of quantum correlations in bipartite quantum systems \cite{Ollivier2001,Henderson2001}. It is defined as the difference between the quantum mutual information and the classical correlations existing in a bipartite system \cite{Yurischev2015,Chakrabarty2011}. For pure bipartite states, the quantum discord coincides with entanglement of formation. The entropic quantum discord of any 2-qubit rank-two state can be calculated exactly. Unfortunately, the situation becomes more complicated for states with rank large than two. To overcome this problem, a geometric variant of quantum discord was introduced to provide an alternative way to deal with non classical correlations in bipartite systems \cite{Dakic2010,Girolami2011}. This geometric measure is defined as Hilbert-Schmidt distance between the considered state $\rho$ and its closest classical state $\chi$. It must be emphasized that the geometric measure of quantum discord by using the Hilbert-Schmidt norm can exhibit less robustness than entanglement in two-qubit systems under special dissipative effects. This unexpected result was reported in \cite{Huang2016}. Furthermore, now it is well established that geometric discord based on Hilbert-Schmidt \cite{Daoud2015,Daoud2014} is not a  faithful measure of quantum correlations \cite{Piani2012}. In fact, this quantifier can increase under local quantum operations acting on the unmeasured qubit.  Other geometric quantifiers have been introduced such as trace distance discord \cite{Paula2013} and trace norm measurement-induced nonlocality \cite{LuoFu2011} to get analytical expressions of quantum discord. A geometric interpretation of one-norm geometric quantum discord for a class of two-qubit X-states is studied in \cite{Huang22016}. Also, the dynamics of entanglement and trace norm measurement-induced nonlocality between two mutually independent atoms interacting with a thermal bath of scalar particles has been investigated in \cite{Huang2018}.\\
Recently, the local quantum uncertainty was introduced by Girolami et al \cite{Girolami2013} (see also the references quoted in \cite{Slaoui2018}) as another quantifier of quantum correlation. This measure is easy to compute analytically for a generic quantum state. It is based on the notion of skew information, introduced by Wigner and Yanase in 1963 \cite{Wigner1963}. The skew information was originally used to describe the information content of mixed states. It plays a fundamental role in quantifying the uncertainty measurements of observables \cite{Luo2005}. It is also related to the concept of Fisher information which is useful in quantum metrology \cite{Luo2003,Frieden2004,Slaoui2019}. Another important feature of the skew information is related to the distinguishability of quantum states \cite{Luo2012}. Moreover, we can use the skew information as a quantifier of quantum coherence \cite{Girolami2014,Du2015,Baumgratz2014}.\\
The aim of this paper is to develop an analytical method to evaluate
quantum correlations in three qubit systems by means of the concept
of local quantum uncertainty. The paper is structured as follows. In section 2, we analyze the quantum correlations in three-qubit $X$ states by employing the concept of local quantum uncertainty. We provide an analytical expression for this quantum correlations quantifier. The obtained results are compared with the tripartite negativity and
entropic quantum discord for some special three-qubit $X$ states. The section 3 is devoted to the dynamics of local quantum uncertainty under decoherence effects and its robustness in some special situations is discussed. In fact, we shall consider the study of the decoherence effects induced by dephasing, depolarizing and phase reversal environments. These different decoherence scenarios are described by employing the Kraus formalism \cite{Kraus1983,Nielson2002}. In section 4, we discuss the monogamy properties of local quantum uncertainty. This is essential to understand the distribution of quantum correlation between the three-qubits family and the whole quantum system. Concluding remarks are given in the last section.
%%%%%%%%%%%%%%%%%%%%%%%%%%%%%%%%%%%%%%%%%%%%%%%%%%%%%%%%%%%%%%%%%%%%%%%%%%%%%%%%%%%%%%%%%%%%%%%
\section{The analytical expression of local quantum uncertainty for three-qubit $X$ states}
\subsection{Definition}
Classically, it is possible to measure any two observables with arbitrary precision. However, for quantum systems, the uncertainty relation imposes a fundamental limit on the precision with which certain pairs of physical properties of a particle can be measured. There are several ways to quantify this uncertainty. In quantum mechanics, the uncertainty of an observable $H$, in a quantum state, is usually quantified by the variance $V\left( {\rho ,H} \right): = {\rm tr}\rho {H^2} - {\left( {\rm tr}\rho H \right)^2}$. However, this relation may exhibits, especially in mixed states, contributions of quantum and classical nature. To deal only with the quantum part of the variance, Wigner and Yanase introduced the notion of skew information. They have shown that the quantum uncertainty relation can be also described in terms of skew information as \cite{Wigner1963}
\begin{equation}\label{skew}
I\left( {\rho ,H} \right): =  - \frac{1}{2}{\rm tr}{\left[\sqrt {{\rho}} ,H \right]^2}.
\end{equation}
This quantity gives the uncertainty of the observable $H$ in the state $\rho$. It reduces to the variance $V\left( {\rho ,H} \right)$ when $\rho$ is a pure state. The skew information is non-negative and vanishes if and only if the density matrix and the observable commute. It is convex. In fact, it does not increase under classical mixing and satisfies the inequality $I\left( {\sum\limits_i {{\lambda _i}{\rho _i},H} } \right) \le \sum\limits_i {{\lambda _i}I\left( {{\rho _i},H} \right)} $ for all quantum states ${{\rho _i}}$ and positive constants ${{\lambda _i}}$ satisfying $\sum\limits_i {{\lambda _i} = 1} $. Based on the concept of skew information, the local quantum uncertainty was introduced recently as a new kind of quantum correlations quantifiers \cite{Girolami2013}. For a qubit-qudit system, the local quantum uncertainty is defined as the minimum skew information achievable by a single local measurement. For a qubit $A$ and a qudit $B$, it writes as \cite{Girolami2013,Slaoui22018}
\begin{equation}
\mathcal{U}\left( {{\rho _{AB}}} \right) = 1 -
{\lambda _{\max }}\left( {{W_{AB}}} \right), \label{lquA}
\end{equation}
where ${\lambda _{\max }}\left( {{W_{AB}}} \right)$ denotes the maximum eigenvalue of the $3 \times 3$ symmetric matrix ${W_{AB}}$ whose elements are given by
\begin{equation}
w_{ij} = {\rm tr}\left\{ {\sqrt {{\rho _{AB}}}  \left( {{\sigma _{i}}
        \otimes {\openone_{d}}} \right)\sqrt {{\rho _{AB}}}  \left( {{\sigma _{j}}
        \otimes {\openone_{{d}}}} \right)} \right\}, \label{w}
\end{equation}
where $\sigma _{i}\left(i=1,2,3\right) $ denote the usual Pauli matrices. Several works were devoted to this type of quantum correlations measure and its dynamics under noisy effects. These works were essentially motivated with the aim to understand how to reduce the decoherence effects and to provide an adequate scheme to protect the quantum correlations for a given decoherence scenario. In this sense, an interesting protocol proposed in \cite{Huang2017} to protect measurement-induced nonlocality and local quantum uncertainty in a two-qubit system passing through an amplitude damping channel. For pure bipartite states $\rho  = \left| {{\psi _{AB}}} \right\rangle \langle {\psi _{AB}}|$, the local quantum uncertainty reduces to the linear entropy of entanglement \cite{Sen2015}
\begin{equation}
{\mathcal{U}}\left( {\left| {{\psi _{AB}}} \right\rangle \left\langle {{\psi _{AB}}} \right|} \right) = 2\left( {1 - {\rm tr}\left( {{\rho _A}^2} \right)} \right).
\end{equation}
\subsection{Derivation of local quantum uncertainty in three qubit $X$ states}
Two qubit states, with non-zero density matrix elements only along the diagonal and anti-diagonal, are called $X$ states because of their visual form resembling the letter $X$. The extension to multi-qubit states was discussed in \cite{Vinjanampathy2011}. In this paper we shall mainly focus on three-qubit states whose density matrices are $X$-shaped. Thus, we consider the family of $X$ states having the following form
\begin{equation}
{\rho _{123}} = \left( {\begin{array}{*{20}{c}}
    {{\rho _{11}}}&0&0&0&0&0&0&{{\rho _{18}}} \\
    0&{{\rho _{22}}}&0&0&0&0&{{\rho _{27}}}&0 \\
    0&0&{{\rho _{33}}}&0&0&{{\rho _{36}}}&0&0 \\
    0&0&0&{{\rho _{44}}}&{{\rho _{45}}}&0&0&0 \\
    0&0&0&{{\rho _{54}}}&{{\rho _{55}}}&0&0&0 \\
    0&0&{{\rho _{63}}}&0&0&{{\rho _{66}}}&0&0 \\
    0&{{\rho _{72}}}&0&0&0&0&{{\rho _{77}}}&0 \\
    {{\rho _{81}}}&0&0&0&0&0&0&{{\rho _{88}}}
    \end{array}} \right), \label{1}
\end{equation}
in the computational basis $\left\{ {\left| {000} \right\rangle ,\left| {010} \right\rangle ,\left| {100} \right\rangle ,\left| {110} \right\rangle ,\left| {001} \right\rangle ,\left| {011} \right\rangle ,\left| {101} \right\rangle ,\left| {111} \right\rangle } \right\}$. The density matrix ${\rho _{123}}$ can be rewritten as
\begin{equation}
{\rho _{123}} = \sum\limits_{i,j=0,1} {{\rho ^{ij}}}  \otimes \left| i \right\rangle \left\langle j \right|, \label{2}
\end{equation}
where the density matrices ${{\rho ^{ij}}}$ are defined by
\begin{equation}
{\rho ^{ii}} = \left( {\begin{array}{*{20}{c}}
    {{\rho _{1 + 4i\hspace{0.05cm} 1 + 4i}}}&{0}&{0}&{0} \\
    {0}&{{\rho _{_{2 + 4i\hspace{0.05cm} 2 + 4i}}}}&{0}&{0} \\
    {0}&{0}&{{\rho _{_{3 + 4i \hspace{0.05cm}3 + 4i}}}}&{0} \\
    {0}&{0}&{0}&{{\rho _{_{4 + 4i\hspace{0.05cm} 4 + 4i}}}}
    \end{array}} \right) \hspace{2cm} i = 1,2, \label{rhoii}
\end{equation}
\begin{equation}
{\rho ^{ij}} = \left( {\begin{array}{*{20}{c}}
    {0}&{0}&{0}&{{\rho _{1 + 4i \hspace{0.05cm}4 + 4j}}} \\
    {0}&{0}&{{\rho _{2 + 4i \hspace{0.05cm}3 + 4j}}}&{0} \\
    {0}&{{\rho _{3 + 4i \hspace{0.05cm}2 + 4j}}}&{0}&{0} \\
    {{\rho _{4 + 4i \hspace{0.05cm} 1 + 4j}}}&{0}&{0}&{0}
    \end{array}} \right) \hspace{2cm} i \ne j. \label{rhoij}
\end{equation}
In the Fano-Bloch representation, the three-qubit state (\ref{1}) writes also as
\begin{equation}
{\rho _{123}} = \frac{1}{8}\sum\limits_{\alpha \beta \gamma } {{R_{\alpha \beta \gamma }}{\sigma _\alpha } \otimes {\sigma _\beta }}  \otimes {\sigma _\gamma },
\end{equation}
where $\alpha $, $\beta $ and $\gamma $ take the values 0, 1, 2 and 3 and the correlation matrix elements ${{R_{\alpha \beta \gamma }}}$ are given by
\begin{equation}
{R_{\alpha \beta \gamma }} = {\rm tr}\left( {{\rho _{123}}\left( {{\sigma _\alpha } \otimes {\sigma _\beta } \otimes {\sigma _\gamma }} \right)} \right). \label{R}
\end{equation}
For states of type (\ref{1}), the non-zero correlation matrix elements ${{R_{\alpha \beta \gamma }}}$, are those with the triplet $(\alpha \beta \gamma$) belonging to the following set
\begin{equation*}
\{(000),(003),(030),(033),(300),(303),(330),(333),
\end{equation*}
\begin{equation*}
(111),(112),(121),(122),(211),(212),(221),(222)\}.
\end{equation*}
The Fano-Bloch representations of the states ${\rho ^{ii}}$ $(i= 1,2)$ (equation (\ref{rhoii})) and ${\rho ^{ij}}$ ($i=1, j=0 $ or $i=0, j=1$) (equation (\ref{rhoij})) are
\begin{equation}
{\rho ^{ii}} = \frac{1}{4}\sum\limits_{\alpha \beta } {R_{_{\alpha \beta }}^{ii}{\sigma _\alpha } \otimes {\sigma _\beta }},
\end{equation}
with $R_{_{\alpha \beta }}^{ii} = {\rm tr}\left( {{\rho ^{ii}}{\sigma _\alpha } \otimes {\sigma _\beta }} \right)$. The non-zero correlation tensor components are given by
\begin{equation}
    \begin{array}{l}
    R_{00}^{ii} = {\rho _{1 + 4i1 + 4i}} + {\rho _{2 + 4i2 + 4i}} + {\rho _{3 + 4i3 + 4i}} + {\rho _{4 + 4i4 + 4i}}\\
    R_{30}^{ii} = {\rho _{1 + 4i1 + 4i}} + {\rho _{2 + 4i2 + 4i}} - {\rho _{3 + 4i3 + 4i}} - {\rho _{4 + 4i4 + 4i}}\\
    R_{03}^{ii} = {\rho _{1 + 4i1 + 4i}} - {\rho _{2 + 4i2 + 4i}} + {\rho _{3 + 4i3 + 4i}} - {\rho _{4 + 4i4 + 4i}}\\
    R_{33}^{ii} = {\rho _{1 + 4i1 + 4i}} - {\rho _{2 + 4i2 + 4i}} - {\rho _{3 + 4i3 + 4i}} + {\rho _{4 + 4i4 + 4i}}.
    \end{array}
\end{equation}
Thus, we have
\begin{equation*}
{\rho ^{ii}} = \frac{1}{4}\left[ {R_{00}^{ii}{\sigma _0} \otimes {\sigma _0} + R_{30}^{ii}{\sigma _3} \otimes {\sigma _0} + R_{03}^{ii}{\sigma _0} \otimes {\sigma _3} + R_{33}^{ii}{\sigma _3} \otimes {\sigma _3}} \right].
\end{equation*}
For the anti-diagonal matrix ${\rho ^{ij}}$ ($i \ne j$), we get
\begin{equation}
{\rho ^{ij}} = \frac{1}{4}\left[ {R_{11}^{ij}{\sigma _1} \otimes {\sigma _1} + R_{12}^{ij}{\sigma _1} \otimes {\sigma _2} + R_{21}^{ij}{\sigma _2} \otimes {\sigma _1} + R_{22}^{ij}{\sigma _2} \otimes {\sigma _2}} \right],
\end{equation}
with
\begin{equation}
R_{11}^{ij} = {\rho _{1 + 4i 4 + 4j}} + {\rho _{4 + 4i1 + 4j}} + {\rho _{2 + 4i 3 + 4j}} + {\rho _{3 + 4i 2 + 4j}},
\end{equation}
\begin{equation}
R_{12}^{ij} = i\left( {{\rho _{1 + 4i4 + 4j}} - {\rho _{4 + 4i1 + 4j}} - {\rho _{2 + 4i3 + 4j}} + {\rho _{3 + 4i2 + 4j}}} \right),
\end{equation}
\begin{equation}
R_{21}^{ij} = i\left( {{\rho _{1 + 4i4 + 4j}} - {\rho _{4 + 4i1 + 4j}} + {\rho _{2 + 4i3 + 4j}} - {\rho _{3 + 4i2 + 4j}}} \right),
\end{equation}
\begin{equation}
R_{22}^{ij} = {\rho _{2 + 4i3 + 4j}} + {\rho _{3 + 4i2 + 4j}} - {\rho _{1 + 4i4 + 4j}} - {\rho _{4 + 4i1 + 4j}}.
\end{equation}
The elements of the tripartite correlations matrix ${R_{\alpha \beta \gamma }}$ (\ref{R}) can be written in terms of bipartite correlation parameters $R_{_{\alpha \beta }}^{ij}$. Indeed, it is simple to see that the density matrix $\rho_{123}$ can be written as
\begin{equation*}
{\rho _{123}} = \frac{1}{2}\left( {{\rho ^{00}} + {\rho ^{11}}} \right) \otimes {\sigma _0} + \frac{1}{2}\left( {{\rho ^{01}} + {\rho ^{10}}} \right) \otimes {\sigma _1} + \frac{i}{2}\left( {{\rho ^{01}} + {\rho ^{10}}} \right) \otimes {\sigma _2} + \frac{1}{2}\left( {{\rho ^{00}} - {\rho ^{11}}} \right) \otimes {\sigma _3}.
\end{equation*}
Furthermore, one verifies that
\begin{equation}
\frac{1}{2}\left( {{\rho ^{00}} + {\rho ^{11}}} \right) = \frac{1}{8}\left[ {R_{00}^{ +  + }{\sigma _0} \otimes {\sigma _0} + R_{03}^{ +  + }{\sigma _0} \otimes {\sigma _3} + R_{30}^{ +  + }{\sigma _3} \otimes {\sigma _0} + R_{33}^{ +  + }{\sigma _3} \otimes {\sigma _3}} \right],
\end{equation}
\begin{equation}
    \frac{1}{2}\left( {{\rho ^{00}} - {\rho ^{11}}} \right) = \frac{1}{8}\left[ {R_{00}^{ -  - }{\sigma _0} \otimes {\sigma _0} + R_{03}^{ -  - }{\sigma _0} \otimes {\sigma _3} + R_{30}^{ -  - }{\sigma _3} \otimes {\sigma _0} + R_{33}^{ -  - }{\sigma _3} \otimes {\sigma _3}} \right],
\end{equation}
\begin{equation}
    \frac{1}{2}\left( {{\rho ^{01}} + {\rho ^{10}}} \right) = \frac{1}{8}\left[ {R_{11}^{ +  - }{\sigma _1} \otimes {\sigma _1} + R_{12}^{ +  - }{\sigma _1} \otimes {\sigma _2} + R_{21}^{ +  - }{\sigma _2} \otimes {\sigma _1} + R_{22}^{ +  - }{\sigma _2} \otimes {\sigma _2}} \right],
\end{equation}
\begin{equation}
    \frac{i}{2}\left( {{\rho ^{01}} - {\rho ^{10}}} \right) = \frac{1}{8}\left[ {R_{11}^{ -  + }{\sigma _1} \otimes {\sigma _1} + R_{12}^{ -  + }{\sigma _1} \otimes {\sigma _2} + R_{21}^{ -  + }{\sigma _2} \otimes {\sigma _1} + R_{22}^{ -  + }{\sigma _2} \otimes {\sigma _2}} \right].
\end{equation}
with
\begin{equation}
    R_{\alpha \beta }^{ +  + } = R_{\alpha \beta }^{00} + R_{\alpha \beta }^{11}, \hspace{1cm} R_{\alpha \beta }^{ -  - } = R_{\alpha \beta }^{00} - R_{\alpha \beta }^{11} \hspace{0.5cm}{\rm for}\hspace{0.5cm} \alpha ,\beta  = 0,3, \label{eq22}
\end{equation}
and
\begin{equation}
    R_{\alpha \beta }^{ +  - } = R_{\alpha \beta }^{01} + R_{\alpha \beta }^{10}, \hspace{1cm} R_{\alpha \beta }^{ -  - } = i\left( {R_{\alpha \beta }^{01} - R_{\alpha \beta }^{10}} \right) \hspace{0.5cm}{\rm for}\hspace{0.5cm} \alpha ,\beta  = 1,2. \label{eq23}
\end{equation}
In this picture, we get
\begin{equation*}
{\rho _{123}} = \frac{1}{8}\left[ {\sum\limits_{\alpha ,\beta  = 0,3} {{R_{\alpha \beta 0}}{\sigma _\alpha } \otimes {\sigma _\beta } \otimes {\sigma _0} + {R_{\alpha \beta 3}}{\sigma _\alpha } \otimes {\sigma _\beta } \otimes {\sigma _3}}  + \sum\limits_{\alpha ,\beta  = 1,2} {{R_{\alpha \beta 1}}{\sigma _\alpha } \otimes {\sigma _\beta } \otimes {\sigma _1} + {R_{\alpha \beta 2}}{\sigma _\alpha } \otimes {\sigma _\beta } \otimes {\sigma _2}} } \right],
\end{equation*}
where the matrix elements $R_{\alpha \beta \gamma}$ are given by
\begin{equation}
    {R_{\alpha \beta 0}} = R_{\alpha \beta }^{ +  + }, \hspace{1cm} {R_{\alpha \beta 3}} = R_{\alpha \beta }^{ -  - } \hspace{0.5cm}{\rm for}\hspace{0.5cm} \alpha ,\beta  = 0,3,
\end{equation}
and
\begin{equation}
    {R_{\alpha \beta 1}} = R_{\alpha \beta }^{ +  - }, \hspace{1cm}{R_{\alpha \beta 2}} = R_{\alpha \beta }^{ -  + } \hspace{0.5cm}{\rm for}\hspace{0.5cm}\alpha ,\beta  = 1,2,
\end{equation}
in term of the correlations matrix elements (\ref{eq22}) and (\ref{eq23}). The three qubit system described by the density matrix (\ref{1}) may be viewed as $2\times4$ quantum systems. The first sub-system is a qubit $(d=2)$ and the second sub-system is a quartet $(d=4)$. In this partitioning scheme, the matrix elements (\ref{w}) write as
\begin{equation}
{w_{ij}} = {\rm tr}\left( {\sqrt {\rho _{123}} \left( {{\sigma _i} \otimes {\sigma _0} \otimes {\sigma _0}} \right)\sqrt {{\rho _{123}}} \left( {{\sigma _j} \otimes {\sigma _0} \otimes {\sigma _0}} \right)} \right),\label{W for X-three}
\end{equation}
where $i$ and $j$ take the values 1, 2, 3. To evaluate $w_{ij}$ (\ref{W for X-three}), we write the matrix $\sqrt {\rho _{123}}$ in the Fano-Bloch representation as $\sqrt {{\rho _{123}}}  = \sum\limits_{\chi \delta \eta } {{T_{\chi \delta \eta }}} {\sigma _\chi } \otimes {\sigma _\delta } \otimes {\sigma _\eta }$. The correlation tensor elements $T_{\chi \delta \eta }$ are given in the appendix. After some algebra, one shown that the matrix elements  (\ref{W for X-three}) are explicitly given by
\begin{equation}
{w_{ij}} = \frac{1}{8}\left[ {\left( {{S_{00}} - \sum\limits_k {{S_{kk}}} } \right){\delta _{ij}} + 2{S_{ij}}} \right], \label{eq27}
\end{equation}
where the quantities $S_{ij}$ are given by
\begin{equation}
S_{00} = \sum\limits_{\chi \delta } {T_{0\chi \delta }{T_{0\chi \delta }}}, \label{S00}
\end{equation}
\begin{equation}
{S_{kk}} = \sum\limits_{\chi \delta } {{T_{k\chi \delta }}{T_{k\chi \delta }}} \hspace{1cm}{\rm for}\hspace{1cm} k = 1,2,3, \label{SKK}
\end{equation}
and
\begin{equation}
{S_{ij}} = \sum\limits_{\chi \delta } {{T_{i\chi \delta }}{T_{j\chi \delta }}}. \label{Sij}
\end{equation}
The matrix elements $w_{ij}$ (\ref{eq27}) can be alternatively expanded as
\begin{equation}
{w_{11}} = \frac{1}{8}\left[ {{S_{00}} + {S_{11}} - {S_{22}} - {S_{33}}} \right]
\end{equation}
\begin{equation}
{w_{22}} = \frac{1}{8}\left[ {{S_{00}} - {S_{11}} + {S_{22}} - {S_{33}}} \right]
\end{equation}
\begin{equation}
{w_{33}} = \frac{1}{8}\left[ {{S_{00}} - {S_{11}} - {S_{22}} + {S_{33}}} \right]
\end{equation}
\begin{equation}
{w_{13}} = {w_{31}} = {w_{23}} = {w_{32}} = 0,
\end{equation}
where the quantities $S_{ij}$ are given by (\ref{S00}), (\ref{SKK}) and (\ref{Sij}). The explicit form of the matrix elements $w_{ij}$ in terms of the density matrix elements is given in the appendix.
\subsection{Illustration}
To exemplify the results obtained above, we consider the special cases of mixed {\rm GHZ} states and three-qubit states of Bell type.
\subsubsection{Mixed {\rm GHZ}-states}
We consider first the three-qubit mixed states of {\rm GHZ} type given by
\begin{equation}
{\rho _{ GHZ}} = \frac{p}{8}\openone_{3} + \left( {1 - p} \right)\left| {\rm GHZ} \right\rangle \left\langle {\rm GHZ} \right|, \label{GHZ}
\end{equation}
where $\left| {\rm GHZ} \right\rangle$ denotes the usual $GHZ$ state: $\left| {\rm GHZ} \right\rangle  = \frac{1}{{\sqrt 2 }}\left( {\left| {000} \right\rangle  + \left| {111} \right\rangle } \right)$. In the computational basis, the state (\ref{GHZ}) takes the form
\begin{equation}
{\rho _{GHZ}} = \frac{1}{8}\left( {\begin{array}{*{20}{c}}
    {4 - 3p}&0&0&0&0&0&0&{4\left( {1 - p} \right)} \\
    0&p&0&0&0&0&0&0 \\
    0&0&p&0&0&0&0&0 \\
    0&0&0&p&0&0&0&0 \\
    0&0&0&0&p&0&0&0 \\
    0&0&0&0&0&p&0&0 \\
    0&0&0&0&0&0&p&0 \\
    {4\left( {1 - p} \right)}&0&0&0&0&0&0&{4 - 3p}
    \end{array}} \right).
\end{equation}
Using the equation (\ref{W11}), (\ref{W22}) and (\ref{W33}), the matrix elements $w_{ij}$ needed to evaluate the local quantum uncertainty, write as
\begin{equation}
{w_{11}} = {w_{22}} = \frac{1}{2}\left( {p + \sqrt {\frac{{p\left( {4 - 3p + \sqrt {p\left( {8 - 7p} \right)} } \right)}}{2}} } \right), \hspace{1cm} {w_{33}} = \frac{{8\sqrt {p\left( {8 - 7p} \right)}  - 31{p^2} + 38p + 1}}{{8\left( {4 - 3p + \sqrt {p\left( {8 - 7p} \right)} } \right)}}.
\end{equation}
The off diagonal matrix elements $w_{ij}$ are zero. It is simple to verify that ${w_{33}} \geqslant {w_{11}}$ for any value of $ p$. This implies that ${\lambda _{\max }}\left( W \right) = {w_{33}}$. Therefore, the local quantum uncertainty is simply given by
\begin{equation}
{\mathcal{U}}\left( {{\rho _{GHZ}}} \right) = \frac{{31{{\left( {p - 1} \right)}^2}}}{{8\left( {4 - 3p + \sqrt {p\left( {8 - 7p} \right)} } \right)}}. \label{lqughz}
\end{equation}
The variation of local quantum uncertainty versus the parameter $p$ is reported in figure 2 and will compared with two other quantum correlations quantifies: the tripartite negativity \cite{Vidal2002} and quantum discord \cite{Giorgi2011}.
\subsubsection{Three-qubit state of Bell type}
As a second instance of three-qubit X-states, we consider states of Bell type \cite{Cabello2002}
\begin{equation}
\rho _B = \frac{1}{8}\left( {{\sigma _0} \otimes {\sigma _0} \otimes {\sigma _0} + \sum\limits_{i = 1}^3 {{c_i}{\sigma _i} \otimes {\sigma _i} \otimes {\sigma _i}} } \right), \label{Bellstate}
\end{equation}
with $0 \le {c_i} \le 1$ and $c_{1}^{2}+c_{2}^{2}+c_{3}^{2}\le 1$. In the computational basis, $\rho _B$ takes the form
\begin{equation}
\rho _B = \frac{1}{8}\left( {\begin{array}{*{20}{c}}
    {1 + {c_3}}&0&0&0&0&0&0&{{c_1} + i{c_2}} \\
    0&{1 - {c_3}}&0&0&0&0&{{c_1} - i{c_2}}&0 \\
    0&0&{1 - {c_3}}&0&0&{{c_1} - i{c_2}}&0&0 \\
    0&0&0&{1 + {c_3}}&{{c_1} + i{c_2}}&0&0&0 \\
    0&0&0&{{c_1} - i{c_2}}&{1 - {c_3}}&0&0&0 \\
    0&0&{{c_1} + i{c_2}}&0&0&{1 + {c_3}}&0&0 \\
    0&{{c_1} + i{c_2}}&0&0&0&0&{1 + {c_3}}&0 \\
    {{c_1} - i{c_2}}&0&0&0&0&0&0&{1 - {c_3}}
    \end{array}} \right).
\end{equation}
In this case also the matrix $W$, whose elements are given by (\ref{W for X-three}), is diagonal. The diagonal elements write
\begin{equation}
{w_{11}} = 1 - \frac{{{c_2}^2 + {c_3}^2}}{{1 + \sqrt {1 - {c_1}^2 - {c_2}^2 - {c_3}^2} }}, \hspace{0.3cm} {w_{22}} = 1 - \frac{{{c_1}^2 + {c_3}^2}}{{1 + \sqrt {1 - {c_1}^2 - {c_2}^2 - {c_3}^2} }}, \hspace{0.3cm} {w_{33}} = 1 - \frac{{{c_1}^2 + {c_2}^2}}{{1 + \sqrt {1 - {c_1}^2 - {c_2}^2 - {c_3}^2} }}.
\end{equation}
Therefore, the local quantum uncertainty in the state ${\rho _B}$ is
given by
\begin{equation}
{\mathcal{U}}\left( {{\rho _B}} \right) = \left\{ \begin{gathered}
\frac{{{c_2}^2 + {c_3}^2}}{{1 + \sqrt {1 - {c_1}^2 - {c_2}^2 - {c_3}^2} }}, \hspace{0.5cm}{\rm if} \hspace{0.5cm} {c_1} \geqslant {c_2}\hspace{0.25cm}{\rm and} \hspace{0.25cm} {c_1} \geqslant {c_3} \hfill \\
\frac{{{c_1}^2 + {c_3}^2}}{{1 + \sqrt {1 - {c_1}^2 - {c_2}^2 - {c_3}^2} }}, \hspace{0.5cm}{\rm if}\hspace{0.5cm}{c_2} \geqslant {c_1}\hspace{0.25cm}{\rm and} \hspace{0.25cm} {c_2} \geqslant {c_3} \hfill \\
\frac{{{c_1}^2 + {c_2}^2}}{{1 + \sqrt {1 - {c_1}^2 - {c_2}^2 - {c_3}^2} }}, \hspace{0.5cm}{\rm if}\hspace{0.5cm}{c_3} \geqslant {c_1}\hspace{0.25cm} {\rm and} \hspace{0.25cm} {c_3} \geqslant {c_2}. \hfill \\
\end{gathered}  \right. \label{lquBell}
\end{equation}
To simplify our numerical analysis and the comparison with other discord-like quantifies, we consider the situation where $c_1 = c_2 = c_3= c$. In this case the equation (\ref{lquBell}) gives
\begin{equation}
{\mathcal{U}}\left( {{\rho _B}} \right) = \frac{{2{c^2}}}{{1 + \sqrt {1 - 3{c^2}} }}.
\end{equation}
The local quantum uncertainty is depicted in Fig 1.
%%%%%%%%%%%%%%%%%%%%%%%%%%%%%%%%%%%%%%%%%%%%%%%%%%%%%%%%%%%%%%%%%%%%%%%%%%%%%%%%%
\subsection{Comparison with quantum discord and negativity}
%%%%%%%%%%%%%%%%%%%%%%%%%%%%%%%%%%%%%%%%%%%%%%%%%%%%%%%%%%%%%%%%%%%%%%%%%%%%%%%%%%%%%%%%%%%%%%%%%%%%%%%%%%%%%ùùù
\subsubsection{Tripartite Negativity}
%%%%%%%%%%%%%%%%%%%%%%%%%%%%%%%%%%%%%%%%%%%%%%%%%%%%%%%%%%%%%%%%%%%%%%%%%%%%%%%%%%%%%%%%%%%%%%%%%%%%%%%%%%%%%ùùù
The entanglement properties of three qubit $X$-states was discussed
in several works (see for instance \cite{Sabin2008,Weinstein2009}).
To decide about entanglement in a tripartite system, the tripartite
negativity ${N^{\left( 3 \right)}}\left( \rho_{123}  \right)$ was
introduced in \cite{Vidal2002}. It is given by
\begin{equation}
{N^{\left(3 \right)}}\left(\rho_{123}\right):= \sqrt[3]{N(\rho_{123}^{T_1})N(\rho_{123}^{T_2})N(\rho_{123}^{T_3})}, \label{eq44}
\end{equation}
where $N\left( {{\rho_{123}^{{T_1}}}} \right)$ denotes the bipartite
negativity between the qubit $1$ and the subsystem $2$ and $3$. It
is defined by $N(\rho_{123}^{T_1})= \sum\limits_i {\left| {{\lambda
_i}\left( {{\rho_{123}^{{T_1}}}} \right)} \right|}  - 1$ where
${{\rho_{123}^{{T_1}}}}$ is the partial transpose of $\rho_{123}$
with respect to the subsystem 1 and ${{\lambda _i}\left(
{{\rho_{123}^{{T_1}}}} \right)}$ are the eigenvalues of
${{\rho_{123}^{{T_1}}}}$. Similar definitions hold for
$N(\rho_{123}^{T_2})$ and $N(\rho_{123}^{T_3})$. The negativity can
be equivalently interpreted as  the most negative eigenvalue of the
partial transpose of the density matrix with respect to qubit $1$.
For tripartite quantum systems with permutation invariance, ${N^{\left( 3\right)}}\left( \rho_{123}\right)$ reduces to the bipartite negativity of any bipartition of the system. This writes as
\begin{equation}
{N^{\left( 3 \right)}}(\rho_{123}) = N(\rho_{123}^{T_1}) = N(\rho_{123}^{T_2}) = N(\rho_{123}^{T_3}). \label{Negativity}
\end{equation}
Thus, for the three-qubit state of Bell type (\ref{Bellstate}), the tripartite negativity vanishes
\begin{equation}
     {N^{\left( 3 \right)}}\left(\rho _B \right) = \frac{1}{2}\left|1 - \sqrt 3 c \right| + \frac{1}{2}\left|1 + \sqrt 3 c \right| - 1=0.
\end{equation}
\begin{figure}[h]
    \centerline{\includegraphics[width=12cm]{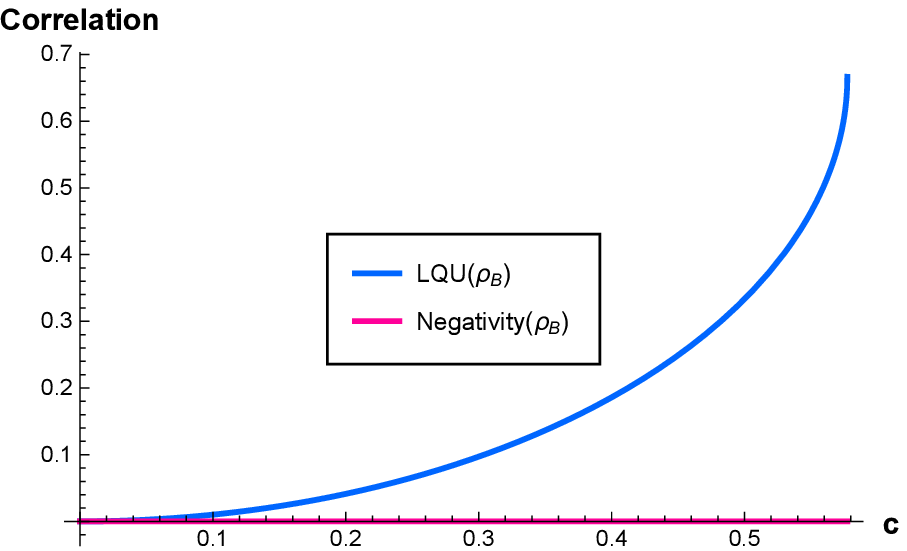}}
    {\bf Figure 1.} {\sf The local quantum uncertainty and negativity in three-qubit state of Bell type versus the parameter $c$.} \label{fg8}
\end{figure}
%%%%%%%%%%%%%%%%%%%%%%%%%%%%%%%%%%%%%%%%%%%%%%%%%%%%%%%%%%%%%%%%%%%%%%%%%%%%%%%%%%%%%%%%%%%%%%%%%%%%%%%ùù
\subsubsection{Tripartite Quantum Discord}
%%%%%%%%%%%%%%%%%%%%%%%%%%%%%%%%%%%%%%%%%%%%%%%%%%%%%%%%%%%%%%%%%%%%%%%%%%%%%%%%%%%%%%%%%%%%%%%%%%%%%%%%%%%%%%%%%%%%%%%%%%ù
According to the reference \cite{Giorgi2011}, the genuine tripartite total correlations $T^{\left( 3\right)}\left( \rho\right) $ in a mixed three-qubit state $\rho_{123}$ is defined by
\begin{equation}
T^{\left( 3\right) }\left( \rho_{123}\right)= T\left( \rho_{123}\right)-T^{\left( 2\right) }\left( \rho_{123}\right), \label{totatcorrelation}
\end{equation}
where $T\left( \rho_{123}\right)= S\left( \rho_{1}\right)+S\left( \rho_{2}\right)+S\left( \rho_{3}\right)-S\left( \rho_{123}\right)$ is the quantum extension of Shannon classical mutual information and $T^{\left( 2\right) }\left( \rho\right)$ is the maximum of the pairwise correlations in the quantum system
\begin{equation}
T^{\left( 2\right) }\left( \rho_{123}\right) ={\rm max}[I^{\left(
2\right) }\left( \rho_{1|2}\right),I^{\left( 2\right) }\left(
\rho_{1|3}\right),I^{\left( 2\right) }\left( \rho_{2|3}\right)]
\end{equation}
with ${I^{\left( 2 \right)}}\left( {{\rho _{i|j}}} \right) = S\left( {{\rho _i}} \right) + S\left( {{\rho _j}} \right) - S\left( {{\rho _{ij}}} \right)$. Here $\rho_{i}\left( i=1,2,3\right) $ is the reduced density matrix for the subsystem $i$ and $S\left( \rho\right) =-{\rm tr}[\rho\log\left( \rho\right) ]$ is the von Neumann entropy. The total tripartite correlation (\ref{totatcorrelation}) rewrites also as
\begin{equation}
    T^{\left( 3\right) }\left(\rho_{123} \right)={\rm min}[I^{\left( 2\right) }\left( \rho_{1|23}\right),I^{\left( 2\right) }\left( \rho_{2|13}\right),I^{\left( 2\right) }\left( \rho_{3|12}\right)].
\end{equation}
Analogously, the genuine tripartite classical correlations $J^{\left( 3\right) }\left( \rho_{123}\right)$ is defined as
\begin{equation}
    J^{\left( 3\right) }\left( \rho_{123}\right)=J\left( \rho_{123}\right) -J^{\left( 2\right)}\left( \rho_{123}\right),
\end{equation}
where
\begin{equation}
J\left( \rho_{123}\right) = \mathop {\max }\limits_{i,j,k \in \left\{ {1,2,3} \right\}} \left[ {S\left( {{\rho _i}} \right) - S\left( {{\rho _{i|j}}} \right) + S\left( {{\rho _k}} \right) - S\left( {{\rho _{k|ij}}} \right)} \right],
\end{equation}
and
\begin{equation}
{J^{\left( 2 \right)}}\left( \rho_{123}\right) = \max \left[ {{J^{\left( 2 \right)}}\left( {{\rho _{1|2}}} \right),{J^{\left( 2 \right)}}\left( {{\rho _{1|3}}} \right),{J^{\left( 2 \right)}}\left( {{\rho _{2|3}}} \right)} \right].
\end{equation}
As for two-qubit systems \cite{Ollivier2001,Henderson2001}, the tripartite quantum discord ${D^{\left( 3 \right)}}\left( \rho_{123}\right)$ can be expressed as the difference between the genuine total correlations and the genuine classical correlations. This is given by
\begin{equation}
    {D^{\left( 3 \right)}}\left( \rho_{123}\right) = {T^{\left( 3 \right)}}\left( \rho_{123}\right) - {J^{\left( 3 \right)}}\left( \rho_{123}\right).
\end{equation}
For three qubit states which are invariant under permutations symmetry, the total and classical correlations can be expressed, respectively, as
\begin{equation}
{T^{\left( 3 \right)}}\left( \rho_{123} \right) = S\left( {{\rho _1}} \right) + S\left( {{\rho _{1|2}}} \right) - S\left( \rho_{123} \right),
\end{equation}
and
\begin{equation}
    {J^{\left( 3 \right)}}\left(\rho_{123}\right) = S\left(\rho _1 \right) - S\left( \rho _{1|23} \right).
\end{equation}
In this case, the tripartite quantum discord reduces to
\begin{equation}
{D^{\left( 3 \right)}}\left(\rho_{123}\right) = S\left(\rho _{1/23} \right) + S\left( {{\rho _{1,2}}} \right) - S\left( \rho  \right), \label{tripartiteDiscord}
\end{equation}
where $S\left( {{\rho _{1|23}}} \right) = \mathop {\min }\limits_{\left\{ {E_{ij}^{23}} \right\}} \sum\limits_{ij} {{p_{ij}}S\left( {{\rho _{1/E_{ij}^{23}}}} \right)} $ is the relative entropy of the qubit $"1"$ when the measurement is carried out on the subsystem (23), the operators ${E_{ij}^{23}}$ are positive-operator-valued measures (POVMs) that act on the qubits $2$ and $3$ (see \cite{Hamieh2004}), ${p_{ij}} = {\rm tr}\left[ {\left( {{\openone^1} \otimes E_{ij}^{23}} \right)\rho } \right]$ is the probability to obtain the $\left(i,j\right) $ outcomes. The density matrix of the system after the measurement is given by ${\rho _{1/E_{ij}^{23}}} = {{{\rm tr}_{23}\left[ {\left( {{\openone^1} \otimes E_{ij}^{23}} \right)\rho } \right]} \mathord{\left/{\vphantom {{T{r_{BC}}\left[ {\left( {{1^A} \otimes E_{ij}^{BC}} \right)\rho } \right]} {{p_{ij}}}}} \right. \kern-\nulldelimiterspace} {{p_{ij}}}}$. For three qubit $X$ states, the quantum discord is given by \cite{Buscemi2013,Beggi2014}
\begin{align}\label{D}
{D^{\left( 3 \right)}}\left( \rho_{123}\right) &= S\left( {{\rho _{1|23}}} \right) - \frac{1}{3}\left( {1 + 4{\rho _{11}}} \right)\log \left( {2 + 8{\rho _{11}}} \right) - \frac{2}{3}\left( {1 - 2{\rho _{11}}} \right)\log \left( {4 - 8{\rho _{11}}} \right) - 1 + 2{\rho _{11}}\log \left( 3 \right) \notag\\ &+ \left( {{\rho _{11}} - {\rho _{18}}} \right)\log \left( {8{\rho _{11}} - 8{\rho _{18}}} \right)
+ \left( {{\rho _{11}} + {\rho _{18}}} \right)\log \left( {8{\rho _{11}} + 8{\rho _{18}}} \right) +\\ & \frac{1}{2}\left( {1 - 2{\rho _{11}} - 6{\rho _{27}}} \right)\log \left( {4 - 8{\rho _{11}} - 24{\rho _{27}}} \right)\notag + \frac{1}{2}\left( {1 - 2{\rho _{11}} + 6{\rho _{27}}} \right)\log \left( {4 - 8{\rho _{11}} + 24{\rho _{27}}} \right),
\end{align}
where the relative entropy $S\left( {{\rho _{1|23}}} \right) = \min \left\{ {{S_1},{S_3}} \right\}$ if $\left| {3{\rho _{18}}} \right| \ge \left| {{\rho _{27}}} \right|$ and ${\rho _{18}}{\rho _{27}}<0$, otherwise it is given by $S\left( {{\rho _{1|23}}} \right) = \min \left\{ {{S_1},{S_2}} \right\}$. The quantities $S_{1}$, $S_{2}$ and $S_{3}$ are given by
\begin{equation}
{S_1} = 1 - \frac{1}{{12}}F\left( 1 - 8{\rho _{11}} \right), \hspace{0.5cm} {S_2} = 1 - \frac{1}{2}G\left( {6{\rho _{27}} + 2{\rho _{18}}} \right), \hspace{0.5cm} {S_3} = 1 - \frac{1}{2}G\left(\sqrt {\frac{{\left( \rho _{18} - {\rho _{27}} \right)^3}}{\rho _{18}}}  \right),  \label{76}
\end{equation}
where the functions $F\left( x \right)$ and $G\left( x \right)$ are defined by
\begin{equation}
    G\left( x \right) = \left( {1 + x} \right)\log \left( {1 + x} \right) + \left( {1 - x} \right)\log \left( {1 - x} \right),
\end{equation}
\begin{equation}
    F\left( x \right) = \left( {3 + x} \right)\log \left( {3 + x} \right) + \left( {3 - 3x} \right)\log \left( {3 - 3x} \right) - 2\left( {3 - x} \right)\log \left( {3 - x} \right).
\end{equation}
Using the expressions (\ref{tripartiteDiscord}), (\ref{D}) and (\ref{76}), the quantum discord in the {\rm GHZ} state (\ref{GHZ}) is given by
\begin{equation}
{D^{\left( 3 \right)}}\left( {{\rho _{GHZ}}} \right) = \left( {\frac{{3p - 4}}{4}} \right)\log \left( {4 - 3p} \right) + \frac{p}{8}\log \left( p \right) + \frac{{8 - 7p}}{8}\log \left( {8 - 7p} \right).
\end{equation}
The amount of quantum correlation is compared, in Fig.2, to the local quantum uncertainty (\ref{lqughz}) and the negativity
\begin{equation}
{N^{\left( 3 \right)}}\left( {{\rho _{GHZ}}} \right) = \left| {\frac{p}{2}} \right| + \left| {\frac{{12 - 9p}}{8}} \right| + \left| {\frac{{5p - 4}}{8}} \right| - 1,
\end{equation}
which is simply obtained from the expressions (\ref{eq44}) and (\ref{Negativity}).
\begin{figure}[H]
    \centerline{\includegraphics[width=9cm]{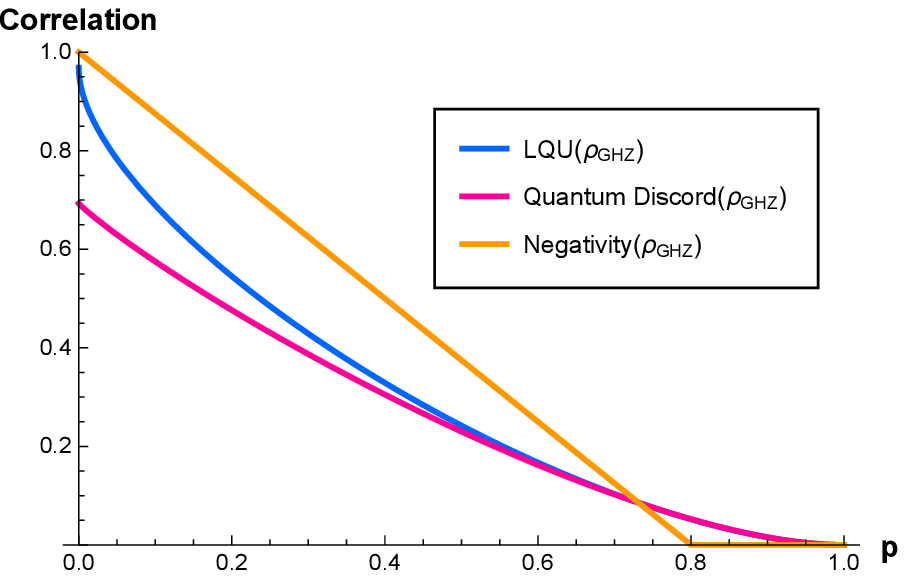}}
    {\bf Figure 2.} {\sf The quantum correlations in tripartite mixed GHZ state measured by local quantum uncertainty, quantum discord and negativity.}
    \label{Fig.1}
\end{figure}
The results plotted in Fig 2 show that local quantum uncertainty and quantum discord present similar behaviour. It is also noticed that like quantum discord, the local quantum uncertainty captures the classical correlations that cannot be revelead by the negativity. Indeed for $0.8 \le p \le 1$, the negativity is zero and cannot be employed as a faithful measure of quantum correlations. Our results agree with the results reported in \cite{Huang2016} for special mixed three-qubit states in which the discord-like measures reveal more quantum correlations than entanglement. Moreover, this shows that local quantum uncertainty constitutes a good quantum correlation quantifier for three-qubit systems.
%%%%%%%%%%%%%%%%%%%%%%%%%%%%%%%%%%%%%%%%%%%%%%%%%%%%%%%%%%%%%%%%%%%%%%%%%%%%%%%%%%%%%%%%%%%%%%%
\section{Local quantum uncertainty in three-qubit {\rm GHZ} state under decoherence channels}
%%%%%%%%%%%%%%%%%%%%%%%%%%%%%%%%%%%%%%%%%%%%%%%%%%%%%%%%%%%%%%%%%%%%%%%%%%%%%%%%%%%%%%%%%%%%%%%%%%%%%ù
The Markov dynamics of states evolving in noisy environments is modeled by the quantum operation $\varepsilon :\rho  \to \varepsilon \left( \rho  \right)$. The channel action on a tripartite state $\rho_{123}$ can be completely characterized in the Kraus representation as follows
\begin{equation}
    \varepsilon \left( \rho_{123} \right) = \sum\limits_{ijk} {\left( {{K_i} \otimes {K_j} \otimes {K_k}} \right)\rho_{123} } {\left( {{K_i} \otimes {K_j} \otimes {K_k}} \right)^\dag },
\end{equation}
where ${{K_i}}$ are the local Kraus operators describing the decoherence of a single qubit. For several decoherence scenarios, the action of the
decoherence channel is generally parameterized by the decoherence probability $q = 1 - \exp \left( { - \kappa t} \right)$ with $\kappa$ is the decay parameter. The Kraus operators satisfy the closure condition ${\sum\limits_i {{K_i}{K_i}} ^\dag } = 1$. In this section we shall discuss the effects of three different environments on the mixed {\rm GHZ}-state (\ref{GHZ}). The dynamics of the quantum correlations is fully characterized by the decoherence parameter $q$.
%%%%%%%%%%%%%%%%%%%%%%%%%%%%%%%%%%%%%%%%%%%%%%%%%%%%%%%%%%%%%%%%%%%%%%%%%%%%%%%%%%%%%%%%%%%%%%%%
\subsection{Dephasing environment}
%%%%%%%%%%%%%%%%%%%%%%%%%%%%%%%%%%%%%%%%%%%%%%%%%%%%%%%%%%%%%%%%%%%%%%%%%%%%%%%%%%%%%%%%%%%%%%%%%%%%ù
We first consider the situation where each qubit is submitted to a dephasing effect induced by the environment. The Kraus operators representing this effect are
\begin{equation}
{K_1} = \left( {\begin{array}{*{20}{c}}
    1&0 \\
    0&{\sqrt {1 - q} }
    \end{array}} \right) \hspace{1cm} {K_2} = \left( {\begin{array}{*{20}{c}}
    0&0 \\
    0&{\sqrt q }
    \end{array}} \right).
\end{equation}
Under this effect, the tripartite state ${\rho _{123}}$ (\ref{1}) remains of $X$-type. The evolved matrix density denoted by $\varepsilon_{DE} \left( \rho_{123} \right)=\rho_{123}^{DE}$ has the same diagonal elements as $\rho_{123}$ while the anti-diagonal elements are multiplied by the factor ${\left( {1 - q} \right)^{\frac{3}{2}}}$. The local quantum uncertainty $\mathcal{U}\left( \rho_{123}^{DE} \right) = 1 -
{\lambda _{\max }}\left( W^{DE} \right)$ can be evaluated using the results reported in the section 2. For the {\rm GHZ}-state (\ref{GHZ}), we shows that the non vanishing matrix elements of the matrix ${W^{DE}}$ (\ref{eq27}), associated with the density matrix $\rho_{123}^{DE}$, are given by
\begin{equation}
w_{11}^{DE} = w_{22}^{DE} = \frac{1}{2}\left( {p + \sqrt {\frac{{p\left( {4 - 3p + \sqrt {{{\left( {4 - 3p} \right)}^2} - 16{{\left( {1 - p} \right)}^2}{{\left( {1 - q} \right)}^3}} } \right)}}{2}} } \right),
\end{equation}
\begin{equation}
w_{33}^{DE} = \frac{{32\left( {1 - p} \right)\left( {1 + \left( {p - 1} \right){{\left( {1 - q} \right)}^3}} \right) + {{\left( {1 + p} \right)}^2} + 4\left( {p + 2\sqrt {{{\left( {4 - 3p} \right)}^2} - 16{{\left( {1 - p} \right)}^2}{{\left( {1 - q} \right)}^3}} } \right)}}{{8\left( {4 - 3p + \sqrt {{{\left( {4 - 3p} \right)}^2} - 16{{\left( {1 - p} \right)}^2}{{\left( {1 - q} \right)}^3}} } \right)}}.
\end{equation}
It is simple to check that $w_{33}^{DE} \ge w_{11}^{DE}$. This gives ${\lambda _{\max }}\left( {{W^{DE}}} \right) = w_{33}^{DE}$ and in this case the local quantum uncertainty is
\begin{equation}
{\mathcal{U}}\left( {\rho _{GHZ}^{DE}} \right) = \frac{{{{\left( {1 - p} \right)}^2}\left( {32{{\left( {1 - q} \right)}^3} - 1} \right)}}{{8\left( {4 - 3p + \sqrt {{{\left( {4 - 3p} \right)}^2} - 16{{\left( {1 - p} \right)}^2}{{\left( {1 - q} \right)}^3}} } \right)}}.
\end{equation}
The local quantum uncertainty is plotted in Fig.3 for different values of mixedness parameter $p$ and the decoherence parameter $q$ which reflects the degradation of quantum correlations under decoherence effects. To investigate the robustness of local quantum uncertainty in comparison with other quantifiers, we analysed also the dynamics of quantum discord and negativity. Using the results (\ref{Negativity}) and (\ref{D}), the negativity in the state ${\rho _{GHZ}^{DE}}$ writes
\begin{equation}
{N^{\left( 3 \right)}}\left( {\rho _{GHZ}^{DE}} \right) = \left| {\frac{p}{2}} \right| + \left| {\frac{{4 - 3p}}{4}} \right| + \left| {\frac{{p - 4\left( {1 - p} \right){{\left( {1 - q} \right)}^{\frac{3}{2}}}}}{8}} \right| + \left| {\frac{{p + 4\left( {1 - p} \right){{\left( {1 - q} \right)}^{\frac{3}{2}}}}}{8}} \right| - 1,
\end{equation}
and the quantum discord (\ref{D}) is given by
\begin{align}
{D^{\left( 3 \right)}}\left( {\rho _{GHZ}^{DE}} \right) &= \frac{{3p - 4}}{4}\log \left( {4 - 3p} \right) + \frac{1}{8}\left( {4 - 3p - 4\left( {1 - p} \right){{\left( {1 - q} \right)}^{\frac{3}{2}}}} \right)\log \left( {4 - 3p - 4\left( {1 - p} \right){{\left( {1 - q} \right)}^{\frac{3}{2}}}} \right){\rm{ }} \notag\\ &+ \frac{1}{8}\left( {4 - 3p + 4\left( {1 - p} \right){{\left( {1 - q} \right)}^{\frac{3}{2}}}} \right)\log \left( {4 - 3p + 4\left( {1 - p} \right){{\left( {1 - q} \right)}^{\frac{3}{2}}}} \right).
\end{align}
\begin{figure}[H]
    \centering
    \begin{minipage}[t]{3in}
        \centering
        \includegraphics[width=3.1in]{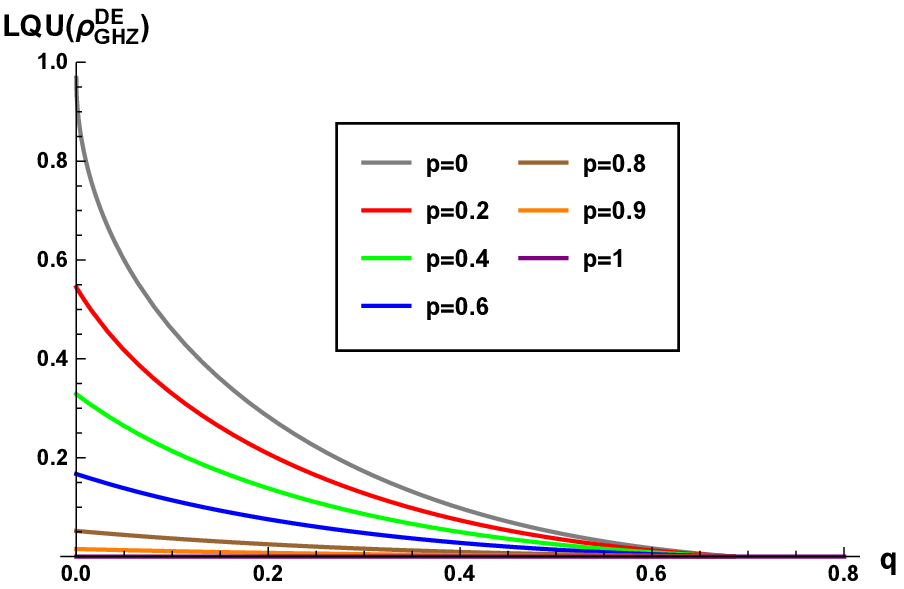}
    \end{minipage}
    \begin{minipage}[t]{3in}
        \centering
        \includegraphics[width=3.1in]{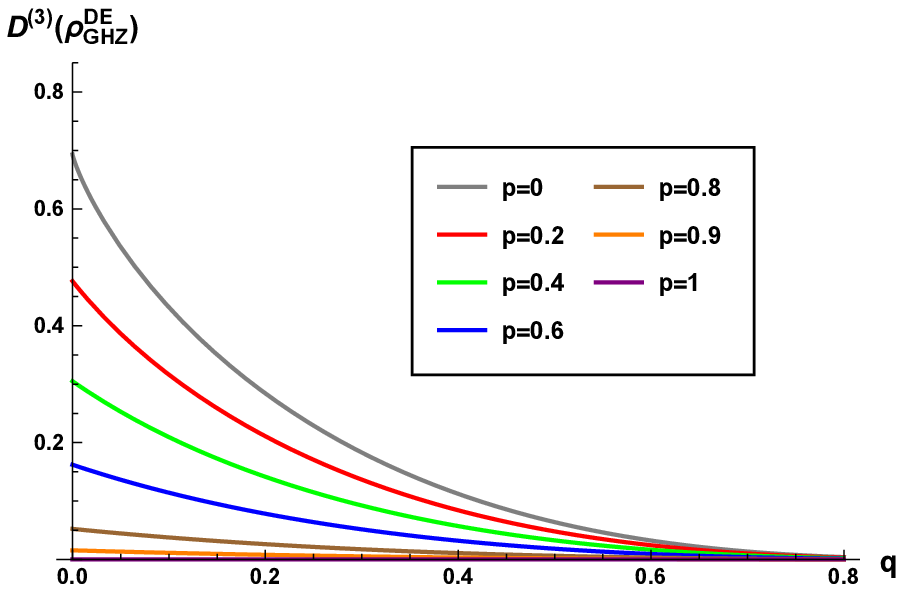}
    \end{minipage}

    {\bf Figure 3.} {\sf The local quantum uncertainty and the quantum discord versus the dephasing parameter $q$. }
\end{figure}
\begin{figure}[H]
    \centerline{\includegraphics[width=9cm]{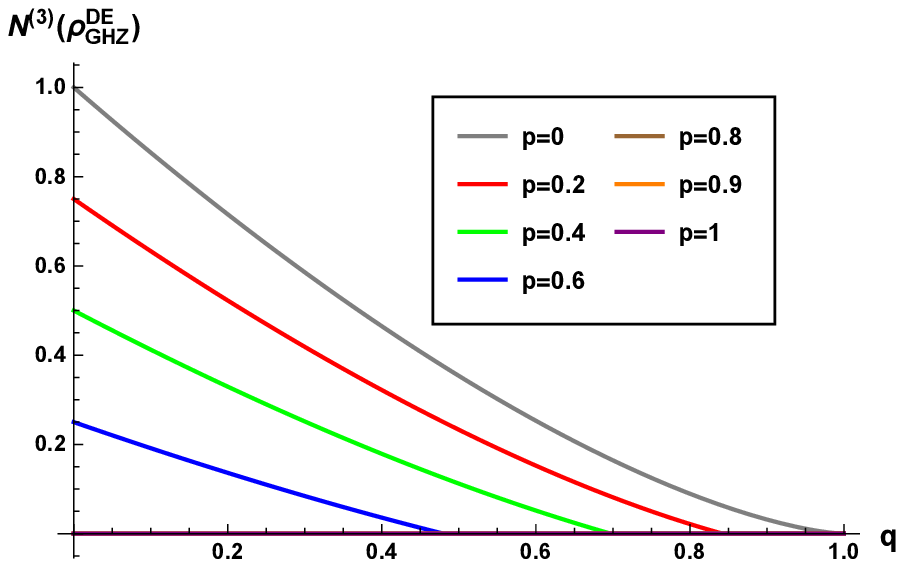}}
    {\bf Figure 4.} {\sf The negativity versus the dephasing parameter $q$.}
    \label{Fig.6}
\end{figure}
The entropic quantum discord in mixed {\rm GHZ} states shows more
robustness against the dephasing effects in comparison with local
quantum uncertainty. On the other hand, the negativity is more
robust than local quantum uncertainty and quantum discord (see
figure 4). Indeed, for $0.8<q<1$, the local quantum uncertainty and
quantum discord vanished whereas the negativity is non-zero for a mixing parameter $p$ taking the values between $0$ and $0.6$. This is an unusual and unexpected important behavior of the negativity in noisy environment in comparison with quantum discord and local quantum uncertainty believed stranger and robust that entanglement.
%%%%%%%%%%%%%%%%%%%%%%%%%%%%%%%%%%%%%%%%%%%%%%%%%%%%%%%%%%%%%%%%%%%%%%%%%%%%%%%%%%%%%%%%%%%%%%%%%%%%
\subsection{Depolarizing environment}
%%%%%%%%%%%%%%%%%%%%%%%%%%%%%%%%%%%%%%%%%%%%%%%%%%%%%%%%%%%%%%%%%%%%%%%%%%%%%%%%%%%%%%%%%%%%%%%%%%%%ù
The depolarization effect of a single qubit in the Kraus representation is given by
\begin{equation}
{K_1} = \sqrt {1 - \frac{{3q}}{4}} \left( {\begin{array}{*{20}{c}}
    1&0 \\
    0&1
    \end{array}} \right), \hspace{0.5cm} {K_2} = \frac{{\sqrt q }}{2}\left( {\begin{array}{*{20}{c}}
    0&1 \\
    1&0
    \end{array}} \right), \hspace{0.5cm} {K_3} = \frac{{\sqrt q }}{2}\left( {\begin{array}{*{20}{c}}
    0&{ - i} \\
    i&0
    \end{array}} \right), \hspace{0.5cm} {K_4} = \frac{{\sqrt q }}{2}\left( {\begin{array}{*{20}{c}}
    1&0 \\
    0&{ - 1}
    \end{array}} \right).
\end{equation}
In this case, the evolved state is also a three-qubit state of $X$ type. The non-zero matrix elements of the density matrix ${\rho _{GHZ}^{PE}}$ are given
\begin{equation}
\rho{'_{11}} = \rho' _{88} = \frac{{4 - 3p}}{8}\left( {1 - \frac{{3q}}{2} + \frac{{3{q^2}}}{4}} \right) + \frac{{3p}}{8}\left( {\frac{q}{2} - \frac{{{q^2}}}{4}} \right),
\end{equation}
\begin{equation}
\rho'_{22} =\rho'_{33} = \rho'_{44} = \rho'_{55} = \rho'_{66} = \rho'_{77} = \frac{p}{8}\left( {1 - \frac{{3q}}{2} + \frac{{3{q^2}}}{4}} \right) + \frac{{4 - p}}{8}\left( {\frac{q}{2} - \frac{{{q^2}}}{4}} \right),
\end{equation}
\begin{equation}
\rho'_{18} = \rho'_{81} = \frac{{\left( {1 - p} \right){{\left( {1 - q} \right)}^3}}}{2},
\end{equation}
in the computational basis. The elements of the matrix (\ref{W for X-three}) write as
{\footnotesize \begin{align}
w_{11}^{PE}= w_{22}^{PE} = &\sqrt {\frac{1}{8} (p + q (1 - p) (2 - q))\left( 4 - 3p + 3q (1 - p) (q - 2) + \sqrt {\left( 4 - 3p + 3q\left( 1 - p \right)\left( {q - 2} \right) \right)^2 - 16{{\left( 1 - p \right)}^2}{\left( 1 - q \right)^6}}  \right)} \notag \\ & + \frac{1}{2}\left( p + q\left(1 - p \right)\left(2 - q \right) \right),
\end{align}}
\begin{align}
    w_{33}^{PE} &= \frac{{{{\left( {1 - p} \right)}^2}{{\left( {1 - q} \right)}^2}\left( {1 - 16{{\left( {1 - q} \right)}^4}} \right)}}{{8\left( {4 - 3p + 3q\left( {1 - p} \right)\left( {q - 2} \right) + \sqrt {{{\left( {4 - 3p + 3q\left( {1 - p} \right)\left( {q - 2} \right)} \right)}^2} - 16{{\left( {1 - p} \right)}^2}{{\left( {1 - q} \right)}^6}} } \right)}} \notag \\ &+ \frac{1}{8}\left( {4 + 3p + 3q\left( {1 - p} \right)\left( {q - 2} \right) + \sqrt {{{\left( {4 - 3p + 3q\left( {1 - p} \right)\left( {q - 2} \right)} \right)}^2} - 16{{\left( {1 - p} \right)}^2}{{\left( {1 - q} \right)}^6}} } \right).
\end{align}
Since $w_{11}^{PE} \leqslant w_{33}^{PE}$, the local quantum uncertainty is given by ${\mathcal{U}}\left( {\rho _{GHZ}^{PE}} \right) = 1 - w_{33}^{PE}$ and one gets
\begin{align}
{\mathcal{U}}\left( {\rho _{GHZ}^{PE}} \right) &= \frac{{{{\left( {1 - p} \right)}^2}{{\left( {1 - q} \right)}^2}\left( {16{{\left( {1 - q} \right)}^4} - 1} \right)}}{{8\left( {4 - 3p + 3q\left( {1 - p} \right)\left( {q - 2} \right) + \sqrt {{{\left( {4 - 3p + 3q\left( {1 - p} \right)\left( {q - 2} \right)} \right)}^2} - 16{{\left( {1 - p} \right)}^2}{{\left( {1 - q} \right)}^6}} } \right)}} \notag \\ &+ \frac{1}{8}\left( {4 - 3p - 3q\left( {1 - p} \right)\left( {2 - q} \right) - \sqrt {{{\left( {4 - 3p + 3q\left( {1 - p} \right)\left( {q - 2} \right)} \right)}^2} - 16{{\left( {1 - p} \right)}^2}{{\left( {1 - q} \right)}^6}} } \right).
\end{align}
The negativity in the evolved state ${\rho _{GHZ}^{PE}}$ is
\begin{align}
    {N^{\left( 3 \right)}}\left( {\rho _{GHZ}^{PE}} \right){\rm{ }} &= \frac{1}{2}\left| {p - 2pq + p{q^2} + 2q - {q^2}} \right| + \frac{1}{4}\left| {4 - 6q + 3{q^2} - 3p + 6pq - 3p{q^2}} \right|- 1 + \\&\notag \frac{1}{8}\left| {p - 2pq + p{q^2} + 2q - {q^2} + 4(1 - p){{(1 - q)}^3}} \right| + \frac{1}{8}\left| {p - 2pq + p{q^2} + 2q - {q^2} - 4(1 - p){{(1 - q)}^3}} \right| ,
\end{align}
and the quantum discord is given by the following expression
{\small\begin{align}
 D^{\left( 3 \right)}\left( \rho _{GHZ}^{PE} \right){\rm{ }} = -& \frac{1}{4}\left( {\left( {4 - 3p} \right)\left( {1 - \frac{{3q}}{2} + \frac{{3{q^2}}}{4}} \right) + 3p\left( {\frac{q}{2} - \frac{{{q^2}}}{4}} \right)} \right)\log \left( {\left( {4 - 3p} \right)\left( {1 - \frac{{3q}}{2} + \frac{{3{q^2}}}{4}} \right) + 3p\left( {\frac{q}{2} - \frac{{{q^2}}}{4}} \right)} \right) \notag \\ & +\frac{1}{8}{\beta _ + }\left( {p,q} \right)\log \left( {{\beta _ + }\left( {p,q} \right)} \right){\rm{  + }} \frac{1}{8}{\beta _ - }\left( {p,q} \right)\log \left( {{\beta _ - }\left( {p,q} \right)} \right),
\end{align}}
where
\begin{equation}
\beta _ \pm \left( {p,q} \right) = \left( {4 - 3p} \right)\left( {1 - \frac{{3q}}{2} + \frac{{3{q^2}}}{4}} \right) + 3p\left( {\frac{q}{2} - \frac{{{q^2}}}{4}} \right) \pm 4\left( {1 - p} \right){\left( {1 - q} \right)^3}.
\end{equation}
\begin{figure}[H]
    \centering
    \begin{minipage}[t]{3in}
        \centering
        \includegraphics[width=3.1in]{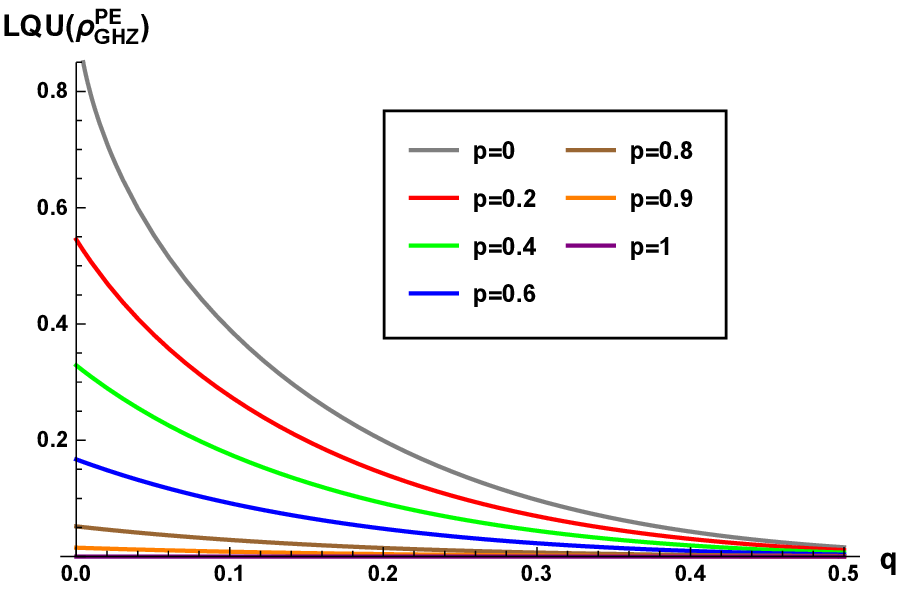}
    \end{minipage}
    \begin{minipage}[t]{3in}
        \centering
        \includegraphics[width=3.1in]{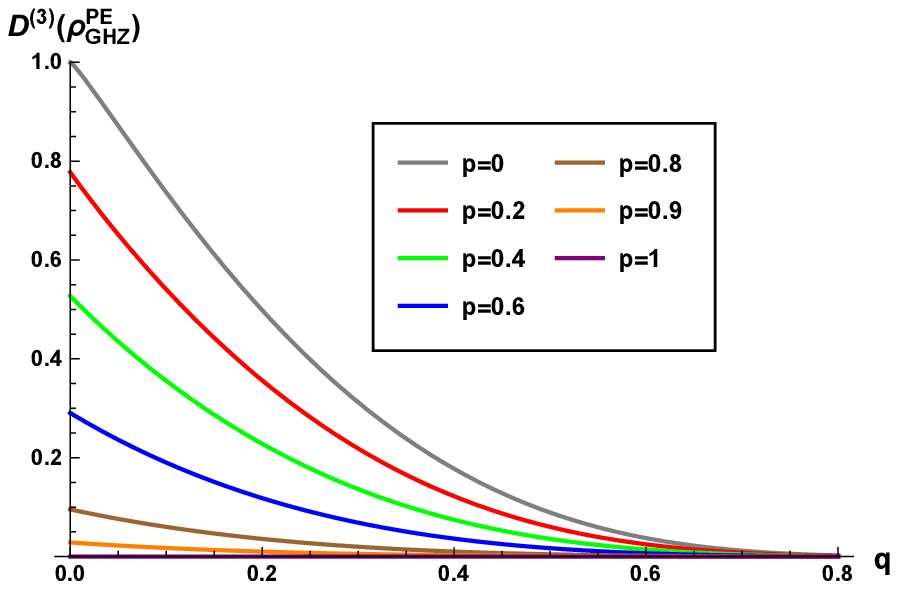}
    \end{minipage}

    {\bf Figure 5.} {\sf The local quantum uncertainty and the quantum discord versus the depolarizing strength $q$. }
\end{figure}
\begin{figure}[H]
    \centerline{\includegraphics[width=9cm]{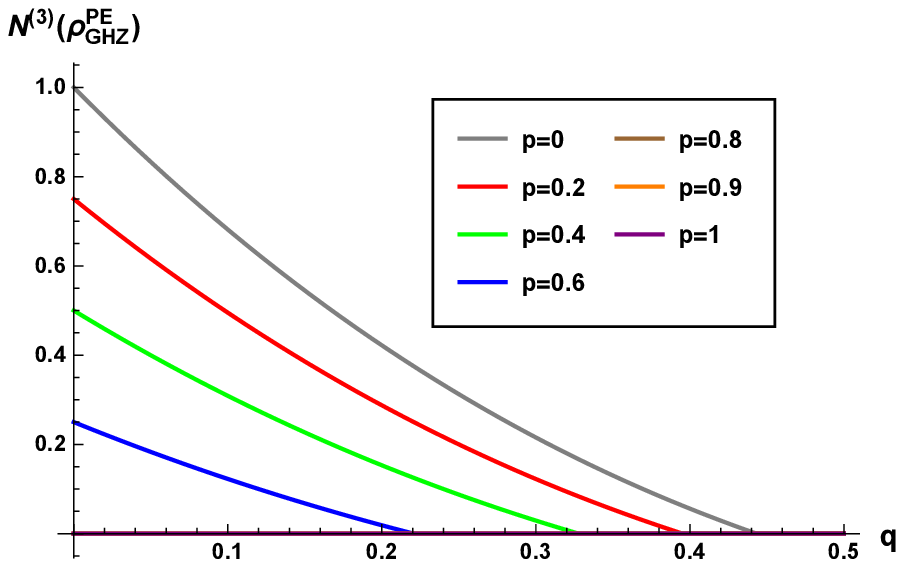}}
    {\bf Figure 6.} {\sf The negativity versus the depolarizing strength $q$.}
\end{figure}
Under depolarizing effects, the quantum discord is more robust in comparison with local quantum uncertainty. However, like for the dephasing effects, it is remarkable that the negativity shows more robustness than the local quantum uncertainty and the quantum discord.
%%%%%%%%%%%%%%%%%%%%%%%%%%%%%%%%%%%%%%%%%%%%%%%%%%%%%%%%%%%%%%%%%%%%%%%%%%%%%%%%%%%%%%%%%%%%%%%%%%%%%%%%%%%%%ù
\subsection{Phase reversal environment:}
%%%%%%%%%%%%%%%%%%%%%%%%%%%%%%%%%%%%%%%%%%%%%%%%%%%%%%%%%%%%%%%%%%%%%%%%%%%%%%%%%%%%%%%%%%%%%%%%%%%ùù
Phase reversal environment leaves the state invariant $\left| 0 \right\rangle$ with the probability $q$ and changes the state $\left| 1 \right\rangle$ to $-\left| 1 \right\rangle$ with the probability $\left( {1 - q} \right)$. The corresponding Kraus operators are
\begin{equation}
{K_1} = \sqrt {1 - q} \left( {\begin{array}{*{20}{c}}
    1&0 \\
    0&1
    \end{array}} \right) \hspace{1cm} {K_2} = \sqrt q \left( {\begin{array}{*{20}{c}}
    1&0 \\
    0&{ - 1}
    \end{array}} \right).
\end{equation}
Under the phase reversal effect, the matrix elements of the evolved density matrix $\rho_{GHZ}^{PRE}$ are $\rho_{ij}^{PRE}=(1 + 10{q^2} - 6q\left( {1 + {q^2}} \right))\rho_{ij}$ for $i \ne j$ and $\rho_{ii}^{PRE}=\rho_{ii}$. In this case, the eigenvalues of the matrix $W^{PRE}$ obtained from the equation (\ref{W for X-three}) are given by
\begin{equation}
w_{11}^{PRE} = w_{22}^{PRE} = \frac{1}{2}\left( {p + \sqrt {\frac{{p\left( {4 - 3p + \sqrt {{{\left( {4 - 3p} \right)}^2} - 16{{\left( {1 - p} \right)}^2}\left( {32\left( {1 + 10{q^2} - 6q\left( {1 + {q^2}} \right)} \right) - 1} \right)} } \right)}}{2}} } \right),
\end{equation}
and
\begin{align}
w_{33}^{PRE} = &\frac{{32\left( {1 - p} \right)\left( {1 + \left( {p - 1} \right)\left( {32\left( {1 + 10{q^2} - 6q\left( {1 + {q^2}} \right)} \right) - 1} \right)} \right) + {{\left( {1 + p} \right)}^2}}}{{8\left( {4 - 3p + \sqrt {{{\left( {4 - 3p} \right)}^2} - 16{{\left( {1 - p} \right)}^2}\left( {32\left( {1 + 10{q^2} - 6q\left( {1 + {q^2}} \right)} \right) - 1} \right)} } \right)}} + \notag \\
&\frac{{\left( {p + 2\sqrt {{{\left( {4 - 3p} \right)}^2} - 16{{\left( {1 - p} \right)}^2}\left( {32\left( {1 + 10{q^2} - 6q\left( {1 + {q^2}} \right)} \right) - 1} \right)} } \right)}}{{2\left( {4 - 3p + \sqrt {{{\left( {4 - 3p} \right)}^2} - 16{{\left( {1 - p} \right)}^2}\left( {32\left( {1 + 10{q^2} - 6q\left( {1 + {q^2}} \right)} \right) - 1} \right)} } \right)}}.
\end{align}
Here also we have $w_{11}^{PRE} \leqslant w_{33}^{PRE}$
and the local quantum uncertainty is simply given by
${\mathcal{U}}\left( {\rho _{GHZ}^{PRE}} \right) = 1 -
w_{33}^{PRE}$. This can be written as
\begin{equation}
{\mathcal{U}}\left( {\rho _{GHZ}^{PRE}} \right) = \frac{{{{\left( {1 - p} \right)}^2}\left( {32\left( {1 + 10{q^2} - 6q\left( {1 + {q^2}} \right)} \right) - 1} \right)}}{{8\left( {4 - 3p + \sqrt {{{\left( {4 - 3p} \right)}^2} - 16{{\left( {1 - p} \right)}^2}\left( {1 + 10{q^2} - 6q\left( {1 + {q^2}} \right)} \right)} } \right)}}.
\end{equation}
For the evolved state $\rho _{GHZ}^{PRE}$, the negativity is given by
\begin{align}
{N^{\left( 3 \right)}}\left( {\rho _{GHZ}^{PRE}} \right) = &\left| {\frac{p}{2}} \right| + \left| {\frac{{4 - 3p}}{4}} \right| + \left| {\frac{{p - 4\left( {1 - p} \right)\left( {1 + 10{q^2} - 6q\left( {1 + {q^2}} \right)} \right)}}{8}} \right|+ \notag\\& \left| {\frac{{p + 4\left( {1 - p} \right)\left( {1 + 10{q^2} - 6q\left( {1 + {q^2}} \right)} \right)}}{8}} \right| - 1,
\end{align}
and the quantum discord writes as
\begin{align}
{D^{\left( 3 \right)}}\left( {\rho _{GHZ}^{PRE}} \right) &=\frac{1}{8}\left( {4 - 3p - 4\left( {1 - p} \right)\left( {1 + 10{q^2} - 6q\left( {1 + {q^2}} \right)} \right)} \right)\log \left( {4 - 3p - 4\left( {1 - p} \right)\left( {1 + 10{q^2} - 6q\left( {1 + {q^2}} \right)} \right)} \right) \notag\\
& +\frac{1}{8}\left( {4 - 3p + 4\left( {1 - p} \right)\left( {1 + 10{q^2} - 6q\left( {1 + {q^2}} \right)} \right)} \right)\log \left( {4 - 3p + 4\left( {1 - p} \right)\left( {1 + 10{q^2} - 6q\left( {1 + {q^2}} \right)} \right)} \right) \notag\\
& +\frac{3p - 4}{4}\log \left( 4 - 3p \right).
\end{align}
\begin{figure}[H]
    \centering
    \begin{minipage}[t]{3in}
        \centering
        \includegraphics[width=3.1in]{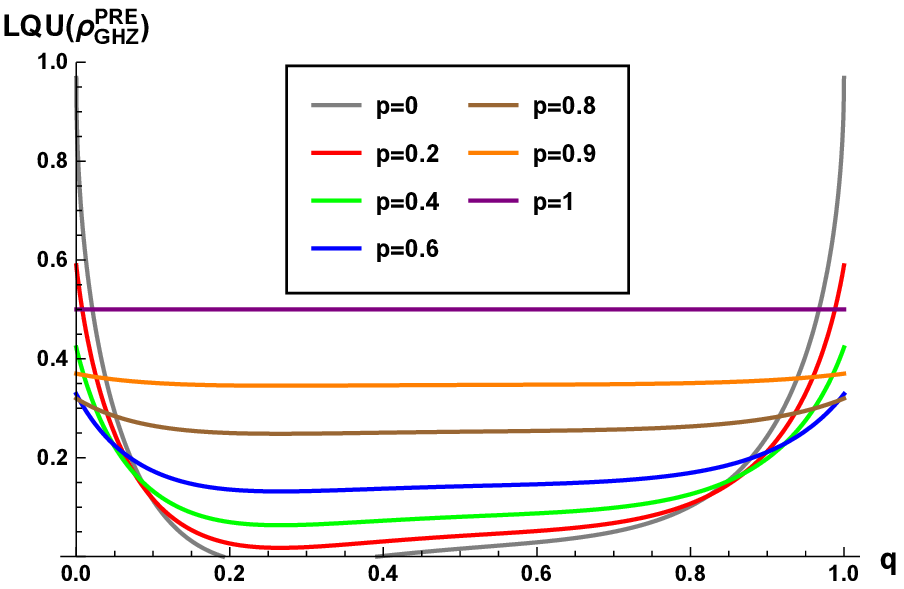}
    \end{minipage}
    \begin{minipage}[t]{3in}
        \centering
        \includegraphics[width=3.1in]{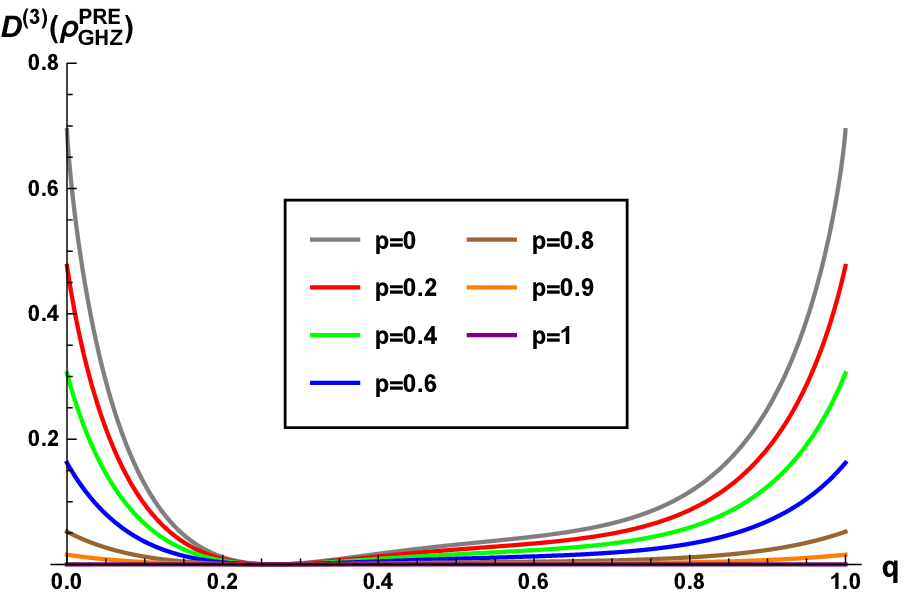}
    \end{minipage}

    {\bf Figure 7.} {\sf The local quantum uncertainty and the quantum discord versus the decoherence parameter $q$. }
\end{figure}

\begin{figure}[H]
    \centerline{\includegraphics[width=9cm]{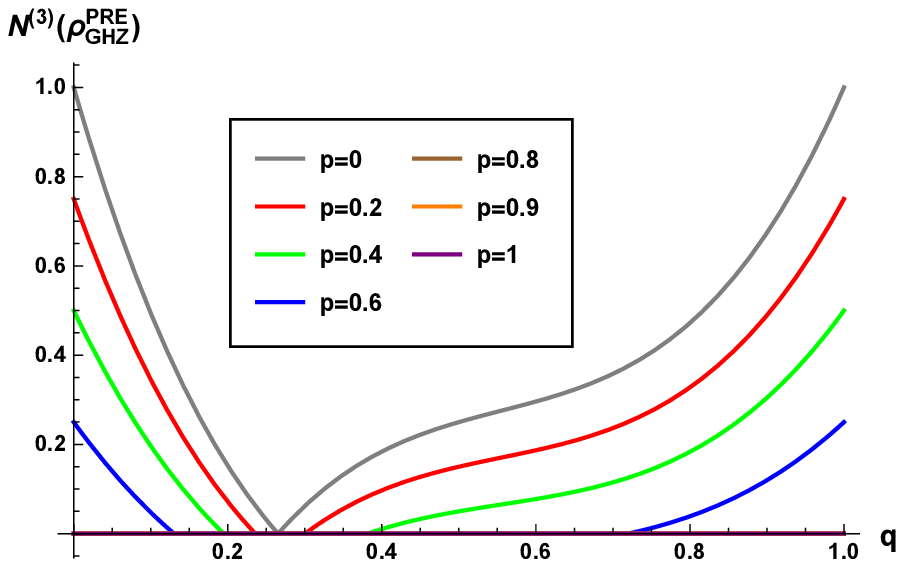}}
    {\bf Figure 8.} {\sf The negativity versus the decoherence parameter $q$.}
\end{figure}
For a mixing parameters $p$ such that $0<p<0.6$, the local quantum
uncertainty shows a revival phenomenon under the phase reversal
environment. It increases to become maximal for higher values of
decoherence parameter $q$. It seems that the phase reversal tends to
enhance the amount of quantum correlation in the system. Also for
states with a mixing parameter $p$ such that $0.7<p<1$, the local
quantum uncertainty tends to be frozen. The entropic quantum discord
does not exhibit this phenomenon. The negativity and entropic
quantum discord confirm the revival of non-classical correlation in
three-qubit {\rm GHZ} under phase reversal effects (see figures 7
and 8).
%%%%%%%%%%%%%%%%%%%%%%%%%%%%%%%%%%%%%%%%%%%%%%%%%%%%%%%%%%%%%%%%%%%%%%%%%%%%%%%%%%%%%%%%%%%%%%%%%%%%%%%%%%%%%
\section{Monogamy of local quantum uncertainty}
%%%%%%%%%%%%%%%%%%%%%%%%%%%%%%%%%%%%%%%%%%%%%%%%%%%%%%%%%%%%%%%%%%%%%%%%%%%%%%%%%%%%%%%%%%%%%%%%%%%%%%%%%
The free shareability relation of classical correlations is no longer valid in the quantum case. Indeed, it has been shown that there are several limitation in sharing quantum correlations between the components of a multi-partite quantum system. These limitations are expressed by the so-called monogamy relation \cite{Kim2012,Streltsov2012}. This limitation was equaled by Coffman, Kundu and Wootters in 2001 for the concurrence and was extended since then to other quantum correlations quantifiers \cite{Coffman2000}. The concept of monogamy can be expressed as follows. Let us denote by $Q_{1|2}$ the shared correlations between the qubits $1$ and $2$, and $Q_{1|3}$ the amount of quantum correlations between the qubits $1$ and $3$. The monogamy constraint imposes that the quantum correlation $Q_{1|23}$ (between the qubit $1$ and the sub-system comprising the qubits $2$ and $3$) is always greater than the sum of $Q_{1|2}$ and $Q_{1|3}$:
\begin{equation} \label{monogamy}
    Q_{1|23}\ge Q_{1|2} + Q_{1|3}.
\end{equation}
Extending this inequality to local quantum uncertainty, the monogamy holds when
\begin{equation}
{{\mathcal{U}}_{1|23}} \ge {{\mathcal{U}}_{1|2}} + {{\mathcal{U}}_{1|3}}.
\end{equation}
We first consider the monogamy of local quantum uncertainty in the mixed {\rm GHZ} state. To do this, we determine the bipartite local quantum uncertainty in the subsystems containing the qubits $2$ and $3$ and the qubits $1$ and $3$. The corresponding reduced density matrices are
\begin{equation}
\rho _{GHZ}^{23} = \rho _{GHZ}^{13} = \frac{1}{4}\left( {\begin{array}{*{20}{c}}
        {2 - p}&0&0&0\\
        0&p&0&0\\
        0&0&p&0\\
        0&0&0&{2 - p}
        \end{array}} \right).
\end{equation}
Using the equation (\ref{lquA}), the elements of the matrix $W$ (\ref{w}) are simply given by
\begin{equation}
    w_{11} = w_{22} = \sqrt {p\left( {2 - p} \right)} \hspace{1cm}{\rm and}\hspace{1cm} w_{33} = 1.
\end{equation}
We have $w_{33} \ge w_{11}$ and the local quantum uncertainty is
\begin{equation}
    {\mathcal{U}}\left( {\rho _{GHZ}^{23}} \right) = {\mathcal{U}}\left( {\rho _{GHZ}^{13}} \right) = 0.
\end{equation}
It is clear that the local quantum uncertainty in the ${{\rho _{GHZ}}}$ satisfies the monogamy constraint
\begin{equation}
{\mathcal{U}}\left( {\rho _{GHZ}} \right) \ge {\mathcal{U}}\left( {\rho _{GHZ}^{23}} \right) + {\mathcal{U}}\left( {\rho _{GHZ}^{13}} \right).
\end{equation}
Similarly, we consider the monogamy property of local quantum uncertainty in the three-qubit Bell states (\ref{Bellstate}). In this case we have $\rho _B^{23} = \rho _B^{13} = \frac{1}{4}{\openone_{4 \times 4}}$ and $w_{11} = w_{22} = {w_{33}} = 1$. Thus, we obtains
\begin{equation}
{\mathcal{U}}\left( {\rho _B^{23}} \right) = {\mathcal{U}}\left( {\rho _B^{13}} \right) = 0,
\end{equation}
and this implies that the local quantum uncertainty in the Bell states (\ref{Bellstate}) is monogamous.
%%%%%%%%%%%%%%%%%%%%%%%%%%%%%%%%%%%%%%%%%%%%%%%%%%%%%%%%%%%%%%%%%%%%%%%%%%%%%%%%%%%%%%%%%%%%%%%%%%%%%%%%%%%%%%
\section{Conclusion}
%%%%%%%%%%%%%%%%%%%%%%%%%%%%%%%%%%%%%%%%%%%%%%%%%%%%%%%%%%%%%%%%%%%%%%%%%%%%%%%%%%%%%%%%%%%%%%%%%%%%%%%%%%%%%

The analytical expression of local quantum uncertainty is derived
for three-qubit $X$ states. As illustration, we computed the non
classical correlation in mixed {\rm GHZ} state and Bell-type
three-qubit state by employing this quantum correlation quantifier.
The obtained results are compared with ones obtained by means of
entropic quantum discord and negativity. The amount of quantum
correlations quantified by local quantum uncertainty is almost
similar to one measured by entropic quantum discord. This indicates
that local quantum uncertainty constitutes an appropriate quantifier
to deal with quantum correlation in multi-qubit systems. This is
essentially due to its easiness computability. Also, it goes beyond
the negativity and offers the tool to quantify the non-classical
correlations contained in multi-qubit separable states. The
evolution of local quantum uncertainty in different noisy
environments is also discussed in this work. Different aspects of
the local quantum uncertainty are compared to the evolution of
entropic quantum discord and negativity. In particular, we have
shown that under phase reversal effect, the local quantum
uncertainty show revival and frozen phenomena. We also investigated
the monogamy property in three qubit state of {\rm GHZ} and Bell
type. We have shown that the monogamy constraint (\ref{monogamy}) is
satisfied. As prolongation of the results obtained
in this work, we believe that it will be interesting to study the
local quantum uncertainty for other three-qubit states, which are
not of $X$-type, as for instance mixed $W$-states of type ${\rho _W}
= \alpha \left| W \right\rangle \left\langle W \right| + \frac{{1 -
\alpha }}{8}{\openone_8}$ where $\alpha  \in {\bf R} $ and
$\left| W \right\rangle  = \frac{1}{{\sqrt 3 }}\left( {\left| {001}
\right\rangle  + \left| {010} \right\rangle  + \left| {100}
\right\rangle } \right)$ denotes a three-qubit $W$-state. We hope to
report on this question in another work.

\section*{Appendix}
In this appendix, we give the necessary tools to compute the local
quantum uncertainty for three-qubit $X$ states. First, the
eigenvalues corresponding to density matrix ${\rho _{123}}$ of the
form (\ref{1}) are given by
\begin{equation}
\begin{gathered}
\lambda _1^ \pm  = \frac{1}{2}{t_1} \pm \frac{1}{2}\sqrt {{t_1}^2 - 4{d_1}},  \hfill \\
\lambda _2^ \pm  = \frac{1}{2}{t_2} \pm \frac{1}{2}\sqrt {{t_2}^2 - 4{d_2}},  \hfill \\
\lambda _3^ \pm  = \frac{1}{2}{t_3} \pm \frac{1}{2}\sqrt {{t_3}^2 - 4{d_3}},  \hfill \\
\lambda _4^ \pm  = \frac{1}{2}{t_4} \pm \frac{1}{2}\sqrt {{t_4}^2 - 4{d_4}},  \hfill \\
\end{gathered}
\end{equation}
where
\begin{equation}
\begin{array}{l}
t_1 = \rho _{11} + \rho _{88},\\
t_2 = \rho _{22} + \rho _{77},\\
t_3 = \rho _{33} + \rho _{66},\\
t_4 = \rho _{44} + \rho _{55},
\end{array} \hspace{2cm} \begin{array}{l}
d_1 = \rho _{11}\rho _{88} - \rho _{18}\rho _{81},\\
d_2 = \rho _{22}\rho _{77} - \rho _{27}\rho _{72},\\
d_3 = \rho _{33}\rho _{66} - \rho _{36}\rho _{63},\\
d_4 = \rho _{44}\rho _{55} - \rho _{45}\rho _{54}.
\end{array}
\end{equation}
The matrix $\sqrt {{\rho _{123}}}$ is also $X$-shaped and has the form
\begin{equation*}
\sqrt {{\rho _{123}}}  = \left( {\begin{array}{*{20}{c}}
    {\frac{{{\rho _{11}} + \sqrt {{d_1}} }}{{\sqrt {{t_1} + 2\sqrt {{d_1}} } }}}&0&0&0&0&0&0&{\frac{{{\rho _{18}}}}{{\sqrt {{t_1} + 2\sqrt {{d_1}} } }}} \\
    0&{\frac{{{\rho _{22}} + \sqrt {{d_2}} }}{{\sqrt {{t_2} + 2\sqrt {{d_2}} } }}}&0&0&0&0&{\frac{{{\rho _{27}}}}{{\sqrt {{t_2} + 2\sqrt {{d_2}} } }}}&0 \\
    0&0&{\frac{{{\rho _{33}} + \sqrt {{d_3}} }}{{\sqrt {{t_3} + 2\sqrt {{d_3}} } }}}&0&0&{\frac{{{\rho _{36}}}}{{\sqrt {{t_3} + 2\sqrt {{d_3}} } }}}&0&0 \\
    0&0&0&{\frac{{{\rho _{44}} + \sqrt {{d_4}} }}{{\sqrt {{t_4} + 2\sqrt {{d_4}} } }}}&{\frac{{{\rho _{45}}}}{{\sqrt {{t_4} + 2\sqrt {{d_4}} } }}}&0&0&0 \\
    0&0&0&{\frac{{{\rho _{54}}}}{{\sqrt {{t_4} + 2\sqrt {{d_4}} } }}}&{\frac{{{\rho _{55}} + \sqrt {{d_4}} }}{{\sqrt {{t_4} + 2\sqrt {{d_4}} } }}}&0&0&0 \\
    0&0&{\frac{{{\rho _{63}}}}{{\sqrt {{t_3} + 2\sqrt {{d_3}} } }}}&0&0&{\frac{{{\rho _{66}} + \sqrt {{d_3}} }}{{\sqrt {{t_3} + 2\sqrt {{d_3}} } }}}&0&0 \\
    0&{\frac{{{\rho _{72}}}}{{\sqrt {{t_2} + 2\sqrt {{d_2}} } }}}&0&0&0&0&{\frac{{{\rho _{77}} + \sqrt {{d_2}} }}{{\sqrt {{t_2} + 2\sqrt {{d_2}} } }}}&0 \\
    {\frac{{{\rho _{81}}}}{{\sqrt {{t_1} + 2\sqrt {{d_1}} } }}}&0&0&0&0&0&0&{\frac{{{\rho _{88}} + \sqrt {{d_1}} }}{{\sqrt {{t_1} + 2\sqrt {{d_1}} } }}}
    \end{array}} \right).
\end{equation*}
In the Fano-Bloch representation, the matrix $\sqrt {{\rho _{123}}}
$ rewrites as
\begin{equation}
\sqrt {{\rho _{123}}}  = \sum\limits_{\chi \delta \eta } {{T_{\chi \delta \eta }}} {\sigma _\chi } \otimes {\sigma _\delta } \otimes {\sigma _\eta },
\end{equation}
where $\chi ,\delta ,\eta  = 0,1,2,3$ and the Fano-Bloch parameters are defined by ${T_{\chi \delta \eta }} = {\rm tr}\left( {\sqrt {{\rho _{123}}} {\sigma _\chi } \otimes {\sigma _\delta } \otimes {\sigma _\eta }} \right)$.\\
The non vanishing elements ${T_{\chi \delta \eta }}$ are given by
\begin{align}
{T_{111}} = &\frac{{{R_{111}} - {R_{221}} - {R_{122}} - {R_{212}}}}{{4\sqrt {{t_1} + 2\sqrt {{d_1}} } }} + \frac{{{R_{111}} - {R_{221}} + {R_{122}} + {R_{212}}}}{{4\sqrt {{t_2} + 2\sqrt {{d_2}} } }} + \\ \notag
& \frac{{{R_{111}} + {R_{221}} + {R_{122}} - {R_{212}}}}{{4\sqrt {{t_3} + 2\sqrt {{d_3}} } }} + \frac{{{R_{111}} + {R_{221}} - {R_{122}} + {R_{212}}}}{{4\sqrt {{t_4} + 2\sqrt {{d_4}} } }},
\end{align}
\begin{align}
{T_{211}} = &\frac{{{R_{112}} + {R_{121}} + {R_{211}} - {R_{222}}}}{{4\sqrt {{t_1} + 2\sqrt {{d_1}} } }} + \frac{{{R_{222}} + {R_{121}} + {R_{211}} - {R_{112}}}}{{4\sqrt {{t_2} + 2\sqrt {{d_2}} } }} + \\ \notag
& \frac{{{R_{121}} - {R_{211}} - {R_{112}} - {R_{222}}}}{{4\sqrt {{t_3} + 2\sqrt {{d_3}} } }} + \frac{{{R_{121}} + {R_{112}} + {R_{222}} - {R_{211}}}}{{4\sqrt {{t_4} + 2\sqrt {{d_4}} } }},
\end{align}
\begin{align}
{T_{121}} =& \frac{{{R_{112}} + {R_{121}} + {R_{211}} - {R_{222}}}}{{4\sqrt {{t_1} + 2\sqrt {{d_1}} } }} + \frac{{{R_{112}} - {R_{222}} - {R_{121}} - {R_{211}}}}{{4\sqrt {{t_2} + 2\sqrt {{d_2}} } }} + \\ \notag
&\frac{{{R_{211}} + {R_{112}} + {R_{222}} - {R_{121}}}}{{4\sqrt {{t_3} + 2\sqrt {{d_3}} } }} + \frac{{{R_{121}} + {R_{112}} + {R_{222}} - {R_{211}}}}{{4\sqrt {{t_4} + 2\sqrt {{d_4}} } }},
\end{align}
\begin{align}
{T_{221}} = &\frac{{{R_{221}} + {R_{122}} + {R_{212}} - {R_{111}}}}{{4\sqrt {{t_1} + 2\sqrt {{d_1}} } }} + \frac{{{R_{111}} - {R_{221}} + {R_{122}} + {R_{212}}}}{{4\sqrt {{t_2} + 2\sqrt {{d_2}} } }} + \\ \notag
&\frac{{{R_{111}} + {R_{221}} + {R_{122}} - {R_{212}}}}{{4\sqrt {{t_3} + 2\sqrt {{d_3}} } }} + \frac{{{R_{122}} - {R_{111}} - {R_{221}} - {R_{212}}}}{{4\sqrt {{t_4} + 2\sqrt {{d_4}} } }},
\end{align}
\begin{align}
{T_{112}} = &\frac{{{R_{112}} + {R_{121}} + {R_{211}} - {R_{222}}}}{{4\sqrt {{t_1} + 2\sqrt {{d_1}} } }} + \frac{{{R_{222}} + {R_{121}} + {R_{211}} - {R_{112}}}}{{4\sqrt {{t_2} + 2\sqrt {{d_2}} } }} + \\ \notag
&\frac{{{R_{211}} + {R_{112}} + {R_{222}} - {R_{121}}}}{{4\sqrt {{t_3} + 2\sqrt {{d_3}} } }} + \frac{{{R_{211}} - {R_{121}} - {R_{112}} - {R_{222}}}}{{4\sqrt {{t_4} + 2\sqrt {{d_4}} } }},
\end{align}
\begin{align}
{T_{122}} = &\frac{{{R_{221}} + {R_{122}} + {R_{212}} - {R_{111}}}}{{4\sqrt {{t_1} + 2\sqrt {{d_1}} } }} + \frac{{{R_{111}} - {R_{221}} + {R_{122}} + {R_{212}}}}{{4\sqrt {{t_2} + 2\sqrt {{d_2}} } }} + \\ \notag
&\frac{{{R_{212}} - {R_{111}} - {R_{221}} - {R_{122}}}}{{4\sqrt {{t_3} + 2\sqrt {{d_3}} } }} + \frac{{{R_{111}} + {R_{221}} - {R_{122}} + {R_{212}}}}{{4\sqrt {{t_4} + 2\sqrt {{d_4}} } }},
\end{align}
\begin{align}
{T_{212}} =& \frac{{{R_{221}} + {R_{122}} + {R_{212}} - {R_{111}}}}{{4\sqrt {{t_1} + 2\sqrt {{d_1}} } }} + \frac{{{R_{221}} - {R_{111}} - {R_{122}} - {R_{212}}}}{{4\sqrt {{t_2} + 2\sqrt {{d_2}} } }} + \\ \notag
&\frac{{{R_{111}} + {R_{221}} + {R_{122}} - {R_{212}}}}{{4\sqrt {{t_3} + 2\sqrt {{d_3}} } }} + \frac{{{R_{111}} + {R_{221}} - {R_{122}} + {R_{212}}}}{{4\sqrt {{t_4} + 2\sqrt {{d_4}} } }},
\end{align}
\begin{align}
{T_{222}} =& \frac{{{R_{222}} - {R_{112}} - {R_{121}} - {R_{211}}}}{{4\sqrt {{t_1} + 2\sqrt {{d_1}} } }} + \frac{{{R_{222}} + {R_{121}} + {R_{211}} - {R_{112}}}}{{4\sqrt {{t_2} + 2\sqrt {{d_2}} } }} +\\ \notag
& \frac{{{R_{211}} + {R_{112}} + {R_{222}} - {R_{121}}}}{{4\sqrt {{t_3} + 2\sqrt {{d_3}} } }} + \frac{{{R_{121}} + {R_{112}} + {R_{222}} - {R_{211}}}}{{4\sqrt {{t_4} + 2\sqrt {{d_4}} } }},
\end{align}
\begin{align}
{T_{000}} = \sqrt {{t_1} + 2\sqrt {{d_1}} }  + \sqrt {{t_2} + 2\sqrt {{d_2}} }  + \sqrt {{t_3} + 2\sqrt {{d_3}} }  + \sqrt {{t_4} + 2\sqrt {{d_4}} },
\end{align}
\begin{align}
{T_{030}} = &\frac{{{R_{030}} + {R_{300}} + {R_{003}} + {R_{333}}}}{{4\sqrt {{t_1} + 2\sqrt {{d_1}} } }} + \frac{{{R_{333}} + {R_{003}} - {R_{030}} - {R_{300}}}}{{4\sqrt {{t_2} + 2\sqrt {{d_2}} } }} + \\ \notag
&\frac{{{R_{300}} + {R_{003}} - {R_{030}} - {R_{333}}}}{{4\sqrt {{t_3} + 2\sqrt {{d_3}} } }} + \frac{{{R_{030}} - {R_{300}} + {R_{003}} - {R_{333}}}}{{4\sqrt {{t_4} + 2\sqrt {{d_4}} } }},
\end{align}
\begin{align}
{T_{300}} = &\frac{{{R_{030}} + {R_{300}} + {R_{003}} + {R_{333}}}}{{4\sqrt {{t_1} + 2\sqrt {{d_1}} } }} - \frac{{{R_{333}} + {R_{003}} - {R_{030}} - {R_{300}}}}{{4\sqrt {{t_2} + 2\sqrt {{d_2}} } }} -\\ \notag
& \frac{{{R_{300}} + {R_{003}} - {R_{030}} - {R_{333}}}}{{4\sqrt {{t_3} + 2\sqrt {{d_3}} } }} + \frac{{{R_{030}} - {R_{300}} + {R_{003}} - {R_{333}}}}{{4\sqrt {{t_4} + 2\sqrt {{d_4}} } }},
\end{align}
\begin{equation}
{T_{330}} = \sqrt {{t_1} + 2\sqrt {{d_1}} }  - \sqrt {{t_2} + 2\sqrt {{d_2}} }  - \sqrt {{t_3} + 2\sqrt {{d_3}} }  + \sqrt {{t_4} + 2\sqrt {{d_4}} },
\end{equation}
\begin{align}
{T_{003}} =& \frac{{{R_{030}} + {R_{300}} + {R_{003}} + {R_{333}}}}{{4\sqrt {{t_1} + 2\sqrt {{d_1}} } }} + \frac{{{R_{030}} + {R_{300}} - {R_{333}} - {R_{003}}}}{{4\sqrt {{t_2} + 2\sqrt {{d_2}} } }} + \\ \notag
&\frac{{{R_{300}} + {R_{003}} - {R_{030}} - {R_{333}}}}{{4\sqrt {{t_3} + 2\sqrt {{d_3}} } }} + \frac{{{R_{300}} + {R_{333}} - {R_{003}} - {R_{030}}}}{{4\sqrt {{t_4} + 2\sqrt {{d_4}} } }},
\end{align}
\begin{equation}
{T_{033}} = \sqrt {{t_1} + 2\sqrt {{d_1}} }  - \sqrt {{t_2} + 2\sqrt {{d_2}} }  + \sqrt {{t_3} + 2\sqrt {{d_3}} }  - \sqrt {{t_4} + 2\sqrt {{d_4}} },
\end{equation}
\begin{equation}
{T_{303}} = \sqrt {{t_1} + 2\sqrt {{d_1}} }  + \sqrt {{t_2} + 2\sqrt {{d_2}} }  - \sqrt {{t_3} + 2\sqrt {{d_3}} }  - \sqrt {{t_4} + 2\sqrt {{d_4}} },
\end{equation}
\begin{align}
{T_{333}} =& \frac{{{R_{030}} + {R_{300}} + {R_{003}} + {R_{333}}}}{{4\sqrt {{t_1} + 2\sqrt {{d_1}} } }} + \frac{{{R_{333}} + {R_{003}} - {R_{030}} - {R_{300}}}}{{4\sqrt {{t_2} + 2\sqrt {{d_2}} } }} + \\ \notag
&\frac{{{R_{030}} + {R_{333}} - {R_{300}} - {R_{003}}}}{{4\sqrt {{t_3} + 2\sqrt {{d_3}} } }} + \frac{{{R_{300}} + {R_{333}} - {R_{003}} - {R_{030}}}}{{4\sqrt {{t_4} + 2\sqrt {{d_4}} } }}.
\end{align}
The eigenvalues $\sqrt {\lambda _1^ \pm } $,$\sqrt {\lambda _2^ \pm
} $,$\sqrt {\lambda _3^ \pm } $ and $\sqrt {\lambda _4^ \pm } $ of
the matrix $\sqrt {{\rho _{123}}} $ are given by
\begin{equation}
\begin{array}{l}
\sqrt {\lambda _1^ \pm }  = \frac{1}{2}\sqrt {{t_1} + 2\sqrt {{d_1}} }  \pm \sqrt {{t_1} - 2\sqrt {{d_1}} }, \\
\sqrt {\lambda _2^ \pm }  = \frac{1}{2}\sqrt {{t_2} + 2\sqrt {{d_2}} }  \pm \sqrt {{t_2} - 2\sqrt {{d_2}} }, \\
\sqrt {\lambda _3^ \pm }  = \frac{1}{2}\sqrt {{t_3} + 2\sqrt {{d_3}} }  \pm \sqrt {{t_3} - 2\sqrt {{d_3}} }, \\
\sqrt {\lambda _4^ \pm }  = \frac{1}{2}\sqrt {{t_4} + 2\sqrt {{d_4}} }  \pm \sqrt {{t_4} - 2\sqrt {{d_4}} }.
\end{array}
\end{equation}
The elements of the matrix W defined by (\ref{w}) are given in terms
of ${R_{\alpha \beta \gamma }}$ and $\sqrt {\lambda _{'i}^ \pm } $
by
\begin{align*}
{w_{12}} = {w_{21}} =& \frac{{\left( {{R_{111}} - {R_{212}}} \right)\left( {{R_{121}} - {R_{222}}} \right) + \left( {{R_{221}} + {R_{122}}} \right)\left( {{R_{112}} + {R_{211}}} \right)}}{{8\left( {\sqrt {\lambda _1^ + }  + \sqrt {\lambda _1^ - } } \right)\left( {\sqrt {\lambda _3^ + }  + \sqrt {\lambda _3^ - } } \right)}} +\\
& \frac{{\left( {{R_{111}} + {R_{212}}} \right)\left( {{R_{121}} + {R_{222}}} \right) + \left( {{R_{122}} - {R_{221}}} \right)\left( {{R_{112}} - {R_{211}}} \right)}}{{8\left( {\sqrt {\lambda _2^ + }  + \sqrt {\lambda _2^ - } } \right)\left( {\sqrt {\lambda _4^ + }  + \sqrt {\lambda _4^ - } } \right)}}
\end{align*}

{\footnotesize{\begin{align*}
        {w_{11}} = &\left( {\sqrt {\lambda _1^ + }  + \sqrt {\lambda _1^ - } } \right)\left( {\sqrt {\lambda _3^ + }  + \sqrt {\lambda _3^ - } } \right) + \left( {\sqrt {\lambda _2^ + }  + \sqrt {\lambda _2^ - } } \right)\left( {\sqrt {\lambda _4^ + }  + \sqrt {\lambda _4^ - } } \right) + \\
        &\frac{{{{\left( {{R_{300}} + {R_{003}}} \right)}^2} + {{\left( {{R_{111}} - {R_{212}}} \right)}^2} - {{\left( {{R_{221}} + {R_{122}}} \right)}^2} + {{\left( {{R_{112}} + {R_{211}}} \right)}^2} - {{\left( {{R_{121}} - {R_{222}}} \right)}^2} - {{\left( {{R_{030}} + {R_{333}}} \right)}^2}}}{{16\left( {\sqrt {\lambda _1^ + }  + \sqrt {\lambda _1^ - } } \right)\left( {\sqrt {\lambda _3^ + }  + \sqrt {\lambda _3^ - } } \right)}} +\\
        & \frac{{{{\left( {{R_{111}} + {R_{212}}} \right)}^2} - {{\left( {{R_{122}} - {R_{221}}} \right)}^2} + {{\left( {{R_{112}} - {R_{211}}} \right)}^2} - {{\left( {{R_{121}} + {R_{222}}} \right)}^2} + {{\left( {{R_{003}} - {R_{300}}} \right)}^2} - {{\left( {{R_{333}} - {R_{030}}} \right)}^2}}}{{16\left( {\sqrt {\lambda _2^ + }  + \sqrt {\lambda _2^ - } } \right)\left( {\sqrt {\lambda _4^ + }  + \sqrt {\lambda _4^ - } } \right)}},
        \end{align*}}}

{\footnotesize{\begin{align*}
        {w_{22}} = &\left( {\sqrt {\lambda _1^ + }  + \sqrt {\lambda _1^ - } } \right)\left( {\sqrt {\lambda _3^ + }  + \sqrt {\lambda _3^ - } } \right) + \left( {\sqrt {\lambda _2^ + }  + \sqrt {\lambda _2^ - } } \right)\left( {\sqrt {\lambda _4^ + }  + \sqrt {\lambda _4^ - } } \right) +\\
        & \frac{{{{\left( {{R_{221}} + {R_{122}}} \right)}^2} - {{\left( {{R_{111}} - {R_{212}}} \right)}^2} + {{\left( {{R_{121}} - {R_{222}}} \right)}^2} - {{\left( {{R_{112}} + {R_{211}}} \right)}^2} + {{\left( {{R_{300}} + {R_{003}}} \right)}^2} - {{\left( {{R_{030}} + {R_{333}}} \right)}^2}}}{{16\left( {\sqrt {\lambda _1^ + }  + \sqrt {\lambda _1^ - } } \right)\left( {\sqrt {\lambda _3^ + }  + \sqrt {\lambda _3^ - } } \right)}} + \\
        &\frac{{{{\left( {{R_{122}} - {R_{221}}} \right)}^2} - {{\left( {{R_{111}} + {R_{212}}} \right)}^2} + {{\left( {{R_{121}} + {R_{222}}} \right)}^2} - {{\left( {{R_{112}} - {R_{211}}} \right)}^2} + {{\left( {{R_{003}} - {R_{300}}} \right)}^2} - {{\left( {{R_{333}} - {R_{030}}} \right)}^2}}}{{16\left( {\sqrt {\lambda _2^ + }  + \sqrt {\lambda _2^ - } } \right)\left( {\sqrt {\lambda _4^ + }  + \sqrt {\lambda _4^ - } } \right)}},
        \end{align*}}}
\begin{align*}
{w_{33}}& =\frac{1}{2}\left[ {{{\left( {\sqrt {\lambda _1^ + }  + \sqrt {\lambda _1^ - } } \right)}^2} + {{\left( {\sqrt {\lambda _3^ + }  + \sqrt {\lambda _3^ - } } \right)}^2} + {{\left( {\sqrt {\lambda _2^ + }  + \sqrt {\lambda _2^ - } } \right)}^2} + {{\left( {\sqrt {\lambda _4^ + }  + \sqrt {\lambda _4^ - } } \right)}^2}} \right] +\\
& \frac{1}{{32}}\left[ {\frac{{{{\left( {{R_{033}} + {R_{300}} + {R_{003}} + {R_{333}}} \right)}^2} - {{\left( {{R_{112}} + {R_{121}} + {R_{211}} - {R_{222}}} \right)}^2} - {{\left( {{R_{111}} - {R_{221}} - {R_{122}} - {R_{212}}} \right)}^2}}}{{{{\left( {\sqrt {\lambda _1^ + }  + \sqrt {\lambda _1^ - } } \right)}^2}}}} \right] +\\
& \frac{1}{{32}}\left[ {\frac{{{{\left( {{R_{333}} + {R_{003}} - {R_{300}} - {R_{030}}} \right)}^2} - {{\left( {{R_{222}} + {R_{121}} + {R_{211}} - {R_{112}}} \right)}^2} - {{\left( {{R_{111}} - {R_{221}} + {R_{122}} + {R_{212}}} \right)}^2}}}{{{{\left( {\sqrt {\lambda _2^ + }  + \sqrt {\lambda _2^ - } } \right)}^2}}}} \right] +\\
& \frac{1}{{32}}\left[ {\frac{{{{\left( {{R_{300}} + {R_{003}} - {R_{030}} - {R_{333}}} \right)}^2} - {{\left( {{R_{121}} - {R_{211}} - {R_{112}} - {R_{222}}} \right)}^2} - {{\left( {{R_{111}} + {R_{221}} + {R_{122}} - {R_{212}}} \right)}^2}}}{{{{\left( {\sqrt {\lambda _3^ + }  + \sqrt {\lambda _3^ - } } \right)}^2}}}} \right] +\\
& \frac{1}{{32}}\left[ {\frac{{{{\left( {{R_{030}} - {R_{300}} + {R_{003}} - {R_{333}}} \right)}^2} - {{\left( {{R_{121}} - {R_{211}} + {R_{112}} + {R_{222}}} \right)}^2} - {{\left( {{R_{111}} + {R_{221}} - {R_{122}} + {R_{212}}} \right)}^2}}}{{{{\left( {\sqrt {\lambda _4^ + }  + \sqrt {\lambda _4^ - } } \right)}^2}}}} \right].
\end{align*}
After some long but feasible calculations, we gets
\begin{equation}
{w_{12}} = {w_{21}} = \frac{{2i\left( {{\rho _{18}}{\rho _{72}} - {\rho _{81}}{\rho _{27}}} \right)}}{{\left( {\sqrt {\lambda _1^ + }  + \sqrt {\lambda _1^ - } } \right)\left( {\sqrt {\lambda _3^ + }  + \sqrt {\lambda _3^ - } } \right)}} + \frac{{2i\left( {{\rho _{36}}{\rho _{54}} - {\rho _{45}}{\rho _{63}}} \right)}}{{\left( {\sqrt {\lambda _2^ + }  + \sqrt {\lambda _2^ - } } \right)\left( {\sqrt {\lambda _4^ + }  + \sqrt {\lambda _4^ - } } \right)}},
\end{equation}
\begin{align}\label{W11}
{w_{11}} = &\left( {\sqrt {\lambda _1^ + }  + \sqrt {\lambda _1^ - } } \right)\left( {\sqrt {\lambda _3^ + }  + \sqrt {\lambda _3^ - } } \right) + \left( {\sqrt {\lambda _2^ + }  + \sqrt {\lambda _2^ - } } \right)\left( {\sqrt {\lambda _4^ + }  + \sqrt {\lambda _4^ - } } \right) + \notag \\
& \frac{{\left( {{\rho _{11}} - {\rho _{88}}} \right)\left( {{\rho _{22}} - {\rho _{77}}} \right) + 2\left( {{\rho _{18}}{\rho _{72}} + {\rho _{81}}{\rho _{27}}} \right)}}{{\left( {\sqrt {\lambda _1^ + }  + \sqrt {\lambda _1^ - } } \right)\left( {\sqrt {\lambda _3^ + }  + \sqrt {\lambda _3^ - } } \right)}} + \frac{{\left( {{\rho _{44}} - {\rho _{55}}} \right)\left( {{\rho _{33}} - {\rho _{66}}} \right) + 2\left( {{\rho _{36}}{\rho _{54}} + {\rho _{45}}{\rho _{63}}} \right)}}{{\left( {\sqrt {\lambda _2^ + }  + \sqrt {\lambda _2^ - } } \right)\left( {\sqrt {\lambda _4^ + }  + \sqrt {\lambda _4^ - } } \right)}},
\end{align}
\begin{align}\label{W22}
{w_{22}} =& \left( {\sqrt {\lambda _1^ + }  + \sqrt {\lambda _1^ - } } \right)\left( {\sqrt {\lambda _3^ + }  + \sqrt {\lambda _3^ - } } \right) + \left( {\sqrt {\lambda _2^ + }  + \sqrt {\lambda _2^ - } } \right)\left( {\sqrt {\lambda _4^ + }  + \sqrt {\lambda _4^ - } } \right) + \notag \\
& \frac{{\left( {{\rho _{11}} - {\rho _{88}}} \right)\left( {{\rho _{22}} - {\rho _{77}}} \right) - 2\left( {{\rho _{18}}{\rho _{72}} + {\rho _{81}}{\rho _{27}}} \right)}}{{\left( {\sqrt {\lambda _1^ + }  + \sqrt {\lambda _1^ - } } \right)\left( {\sqrt {\lambda _3^ + }  + \sqrt {\lambda _3^ - } } \right)}} + \frac{{\left( {{\rho _{44}} - {\rho _{55}}} \right)\left( {{\rho _{33}} - {\rho _{66}}} \right) - 2\left( {{\rho _{36}}{\rho _{54}} + {\rho _{45}}{\rho _{63}}} \right)}}{{\left( {\sqrt {\lambda _2^ + }  + \sqrt {\lambda _2^ - } } \right)\left( {\sqrt {\lambda _4^ + }  + \sqrt {\lambda _4^ - } } \right)}},
\end{align}
\begin{align}\label{W33}
{w_{33}} = &\frac{1}{2}\left( {1 + 2\sum\limits_{i = 1}^4 {\sqrt {{d_i}} } } \right) + \frac{{{{\left( {2{\rho _{11}} + {\rho _{66}} - {\rho _{88}} - {\rho _{77}} - {\rho _{55}}} \right)}^2} - 16{\rho _{18}}{\rho _{81}}}}{{8{{\left( {\sqrt {\lambda _1^ + }  + \sqrt {\lambda _1^ - } } \right)}^2}}} + \frac{{{{\left( {{\rho _{33}} - {\rho _{66}}} \right)}^2} - 4{\rho _{36}}{\rho _{63}}}}{{2{{\left( {\sqrt {\lambda _4^ + }  + \sqrt {\lambda _4^ - } } \right)}^2}}} \notag \\
& + \frac{{{{\left( {{\rho _{44}} - {\rho _{55}}} \right)}^2} + {{\left( {{\rho _{63}} - {\rho _{36}}} \right)}^2} - {{\left( {{\rho _{45}} + {\rho _{54}}} \right)}^2}}}{{2{{\left( {\sqrt {\lambda _2^ + }  + \sqrt {\lambda _2^ - } } \right)}^2}}} + \frac{{{{\left( {{\rho _{22}} - {\rho _{77}}} \right)}^2} - 4{\rho _{27}}{\rho _{72}}}}{{2{{\left( {\sqrt {\lambda _3^ + }  + \sqrt {\lambda _3^ - } } \right)}^2}}}.
\end{align}
with $\lambda^{\pm} (i=1,2, 3, 4)$ are the eigenvalues of the density matrix $\rho_{123}$ (\ref{1}).

\end{document}